\documentclass[pra,aps,superscriptaddress,twocolumn,showpacs]{revtex4-1}
\usepackage{graphicx,amsfonts,bm,amsmath}
\newcommand{\zt}{\tilde z}
\newcommand{\tit}{\tilde t}

\newcommand{\omt}{\tilde \omega}

\begin{document}

\author{Emil Zeuthen}
\thanks{These authors contributed equally to this work}
%\email{zeuthen@nbi.dk}
\affiliation{QUANTOP, Danish National Research Foundation Center for Quantum Optics,
Niels Bohr Institute, University of Copenhagen, DK-2100 Copenhagen \O, Denmark}
\author{Anna Grodecka-Grad}
\thanks{These authors contributed equally to this work}
\email{anna.grodecka-grad@nbi.dk}
\affiliation{QUANTOP, Danish National Research Foundation Center for Quantum Optics,
Niels Bohr Institute, University of Copenhagen, DK-2100 Copenhagen \O, Denmark}
\author{Anders S. S{\o}rensen}
%\email{anders.sorensen@nbi.dk}
\affiliation{QUANTOP, Danish National Research Foundation Center for Quantum Optics,
Niels Bohr Institute, University of Copenhagen, DK-2100 Copenhagen \O, Denmark}

\title{Three-dimensional theory of quantum memories based on $\Lambda$-type atomic ensembles}

\begin{abstract}
We develop a three-dimensional theory for quantum memories based on light storage
in ensembles of $\Lambda$-type atoms, where two long-lived atomic ground states are employed. 
We consider light storage in an ensemble of finite spatial extent and we show that within the paraxial approximation 
the Fresnel number of the atomic ensemble and the optical depth
are the only important physical parameters determining the quality of the quantum memory.
We analyze the influence of these parameters on the storage of light 
followed by either forward or backward read-out from the quantum memory. 
We show that for small Fresnel numbers, the forward memory provides higher efficiencies, whereas for large Fresnel numbers,
the backward memory is advantageous. 
The optimal light modes to store in the memory are presented together with the corresponding spin-waves and outcoming light modes.
We show that for high optical depths such $\Lambda$-type atomic ensembles allow for highly efficient backward and forward 
memories even for small Fresnel numbers~$F\gtrsim 0.1$. 
\end{abstract}

\maketitle

%%%%%%%%%%%%%%%%%%%%%%%%%%%%%%%%%%%%%%%%%%
\section{Introduction}
%%%%%%%%%%%%%%%%%%%%%%%%%%%%%%%%%%%%%%%%%%

The natural way to transmit quantum information is to use photons. A photonic state, however, has to be stored locally 
in order to process the information or to use it at a later time. To this end, one needs an efficient and controllable
quantum interface between light and matter that will store incoming light as a stationary excitation
and release it at a later time preserving quantum correlations.
This can be realized by, for instance, an ensemble of atoms or impurities in a solid state host, 
see Ref.~\cite{Hammerer2010} for an extensive review. 
The resulting quantum memory is an essential part of many quantum information devices.
It can allow for long distance quantum communication~\cite{Sangouard2011} and, in connection with generation of entanglement 
and its swapping~\cite{duan2001}, it might be a basic building block for a quantum internet~\cite{kimble2008}.

One way to achieve strong coupling between light and matter is to use a single atom in a cavity \cite{Gorshkov2007,Boozer2007,Specht2011}.
This is, however, hard to realize experimentally and allows only for storing a single mode. 
An alternative approach is to use atomic ensembles, where the coupling strength scales with
the number of atoms \cite{Hammerer2010}.  Its scalability allows for multimode memories \cite{Sorensen2008,Nunn2008} 
and it is easier to implement in practice. 
Quantum memories have already been demonstrated in a number of experiments based on atomic ensembles,
see e.g. \cite{Julsgaard2004,Novikova2007,Vudyasetu2008,Firstenberg2009,Choi2010,Reim2010,Hosseini2011}
as well as solid state systems, see e.g. \cite{Afzelius2010,Bonarota2011,Saglamyurek2011}.
Despite the fast experimental progress in the field of single and multimode quantum memories, 
little work has been done on describing the full spatial profile of the 
excitations~\cite{Sorensen2008,Sorensen2009,Porras2008,Pedersen2009}.
The widely used one-dimensional theory for spatial quantum memories assumes only a single 
transverse mode \cite{Gorshkov2007a,Gorshkov2007c}. It has been shown that this theory 
is correct for infinitely large atomic ensembles~\cite{Hammerer2010,Sorensen2008,Sorensen2009}.

Within the one-dimensional approach \cite{Gorshkov2007a,Gorshkov2007c} it has been demonstrated 
that the optimal control strategy for storage and retrieval of light can be applied to a  wide class of quantum memory
schemes including electromagnetically induced transparency (EIT) \cite{Liu2001}, off-resonant Raman \cite{Kozhekin2000}, 
and photon-echo based \cite{Moiseev2001} methods. All of these schemes yield identical maximum efficiencies.
Moreover, the maximal efficiency depends only on one physical parameter: the optical depth of the medium $d$.
The retrieval of the light stored in the atomic spin-wave can be interpreted in terms of constructive interference of light emission
in the forward direction set by the direction of the collinear control beam giving an effective rate $\gamma d$.
Thus the efficiency is given by the ratio between this emission and the unwanted one:
\begin{equation}
\eta \sim \frac{\gamma d}{\gamma d + \gamma} \sim 1-\frac{1}{d}.
\end{equation}
The efficiency is independent of the detuning and the temporal shape of the control field
and only depends on the shape of the stored quantum light pulse and the optical depth $d$.
Furthermore, the ideal storage process can be seen as the time reversed of the read-out, thus also depending only on the optical depth $d$.

In this paper, we develop a three-dimensional theory of quantum memory for light based on $\Lambda$-type atomic ensembles, 
where two long-lived atomic ground states are used.
We solve the problem of light storage in an atomic ensemble of finite spatial extent,
where the spatial distribution of the stored stationary excitation (the spin-wave) depends
on the transverse profile and temporal shape of the input light pulse. 
It is shown that within paraxial approximation there are only two physical parameters that determine the quality of the memory:
the optical depth $d$ and the Fresnel number of the atomic ensemble~$F$,
whereas in the one-dimensional theory, where $F\rightarrow \infty$, only the optical depth matters.
The same conclusion was reached in Ref.~\cite{Sorensen2009}, where a three-dimensional theory of
the closely related problem of stimulated Raman
scattering was developed. There, semi-analytical expressions for total emitted light were derived,
whereas in this paper we perform exact numerical calculations describing the full process of storage
followed by retrieval of light. In addition, we optimize this combined process to obtain the best performance
of the spatial quantum memory and its dependence on the physical parameters $F$ and~$d$.
We calculate the efficiencies of storage followed by either forward or backward retrieval of light
and study their dependence on the two crucial parameters of the atomic ensemble.
We find that the direction which has the highest read-out efficiency depends on the Fresnel number of the ensemble. 
The optimal spin-wave modes with highest achievable efficiencies of the combined process of storage followed by retrieval from 
the memory are presented together with the optimal incoming and outgoing light modes.
In addition, we show that the optimal input light pulses are well approximated by the product of a temporal and transverse spatial shape. 
This is advantageous from an experimental point of view, since one does not need to vary the transverse profile in time. 
We show that for time independent driving the optimal output light modes are the complex conjugates of the time reversed input modes. 
In addition, we calculate the relevant experimental parameters of the optimal input light modes, that is the beam waist 
and the distance between the focal plane and the beginning of the ensemble. 
As for the one-dimensional case, a large class of resonant (EIT) and off-resonant Raman based experiments
are contained in the presented three-dimensional theory. 
We show that for optically dense atomic ensembles, the studied quantum memory leads to high efficiencies
of storage followed by backward or forward read-out even for small Fresnel numbers of the sample ($F \gtrsim 0.1$).

The remainder of the paper is organized as follows:
In Sec.~\ref{sec:model}, the model of the quantum memory is presented. 
Next, in Sec.~\ref{sec:read}, we analyze the read-out process and its dependence on the crucial physical parameters:
the optical depth and the Fresnel number of the ensemble. 
The adiabatic storage and retrieval is studied in Sec ~\ref{sec:full},
where the analytical expressions are derived in Sec.~\ref{sec:forw}.
The results of numerical calculations for both read-out directions are presented in Sec.~\ref{sec:results}
together with the discussion of the implications of the three-dimensional theory. 
Sec.~\ref{sec:concl} concludes the paper with final remarks.
In addition, some technical details and further analysis are presented in the Appendixes.

%%%%%%%%%%%%%%%%%%%%%%%%%%%%%%%%%%%%%%%%%%
\section{Quantum memory model}\label{sec:model}
%%%%%%%%%%%%%%%%%%%%%%%%%%%%%%%%%%%%%%%%%%

We consider an atomic ensemble of finite spatial extent which contains atoms with a $\Lambda$-type level structure [see Fig.~1 (left)].
The weak quantum field carries the information to be stored and couples states $|0\rangle$ and $|e\rangle$ with coupling strength $g$.
The strong classical field couples the states $|1\rangle$ and $|e\rangle$ with Rabi frequency $\Omega(t)$ and controls 
the storage and retrieval of light into and from the memory.
The storage is performed as follows: from each excitation removed from the quantum light field one atomic excitation goes from
$|0\rangle$ to $|1\rangle$ via the excited state~$|e\rangle$ and the information is stored in a collective state of the ensemble
spin-wave excitations $S \sim \sum_i |0 \rangle_i\langle 1|$.

\begin{figure}[t]
\begin{center} 
\unitlength 1mm
{\resizebox{42mm}{!}{\includegraphics{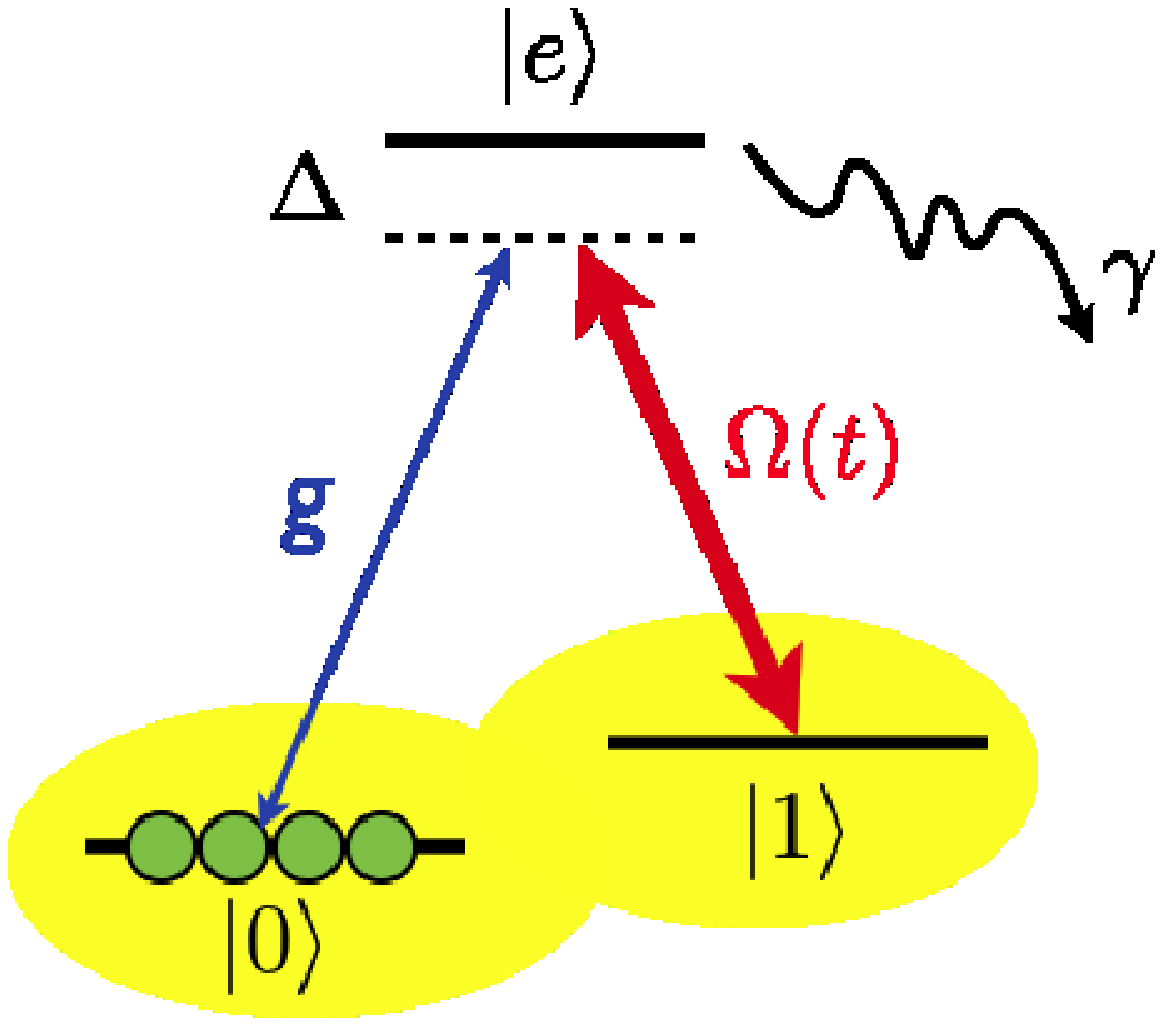}}}
{\resizebox{43mm}{!}{\includegraphics{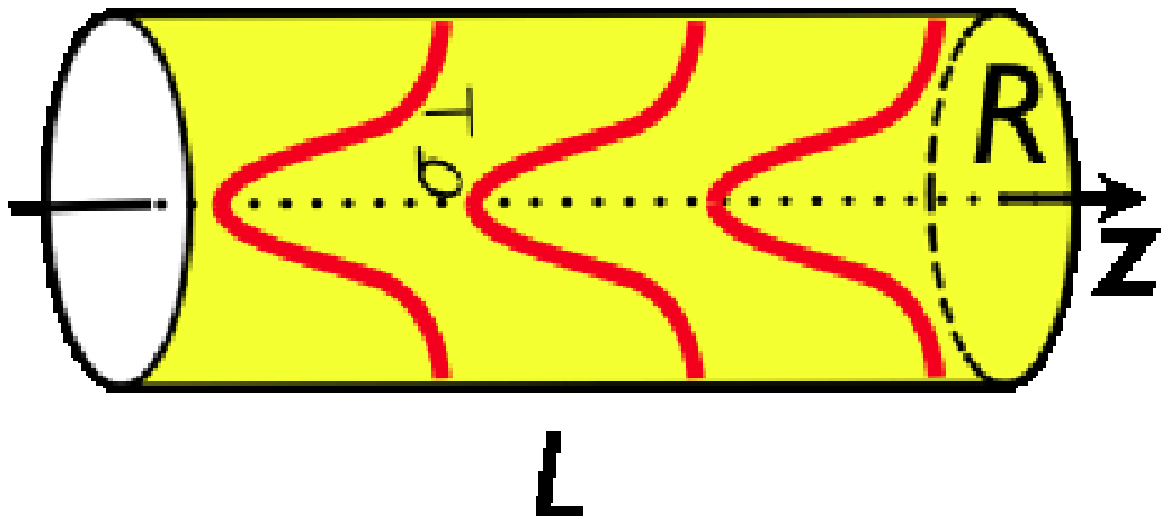}}}
\end{center} 
\caption{\label{fig:scheme}(Color online) (left) A schematic plot of the level scheme of the considered $\Lambda$-type atoms
with two long-lived ground states $|0\rangle$ and $|1\rangle$. The quantum field carries the information to be stored 
and couples $|0\rangle$ and $|e\rangle$ with coupling strength $g$. States $|1\rangle$ and $|e\rangle$
are coupled by the strong classical control field with Rabi frequency $\Omega(t)$. $\Delta$ denotes the detuning of the light
from the excited state $|e\rangle$, which can spontaneously decay at a rate $\gamma$.
(right) Atomic ensemble of cylindrical symmetry with Gaussian distribution of atoms in the radial direction.
The density along the $z$ direction is constant. The length of the sample is $L$ and the width is characterized by~$\sigma_\perp$.}
\end{figure}

We assume an atomic ensemble with cylindrically symmetric density distribution [see Fig.~\ref{fig:scheme} (right)]
and the solution strategies presented below will work for any such atomic density distribution. 
For concreteness, however, we choose a Gaussian distribution in the radial direction 
with $n(\vec r) = n_0 \exp \left({-\frac{\rho^2}{2\sigma_\perp^2}}\right)$,
where the normalization factor is $n_0 = N_{\rm A}/(2\pi L \sigma_\perp^2)$
with $N_{\rm A} = \int d^{3}\vec r n(\vec r)$ the number of atoms,
and $\sigma_\perp$ the width of the ensemble ($\sigma_\perp^2$ being the corresponding cross section area).
For simplicity, we assume a constant density along the $z$ axis,~$n_0$ for $z\in[0,L]$,
so that our ensemble forms a cylinder of length $L$.

To solve the three-dimensional problem of storage of a set of light modes, 
we expand the electric field into a complete and orthogonal set of modes $\vec f_{k,mn}(\vec r)$,
which are solutions to the Helmholtz equation. For simplicity, we employ the paraxial approximation here
and assume that the polarization $\vec e_k$ factors out $\vec f_{k,mn}(\vec r) = \vec e_k f_{k,mn}(\vec r)$. 
The light beams under consideration are cylindrically symmetric, collinear, 
and propagate in the $z$ direction and are conveniently described in a basis set of Bessel beams:
\begin{eqnarray}\label{eq:ff}
f_{k,mn}(\rho,\phi,z) & = & N_{mn} e^{im\phi} J_{m}(k_{\perp,mn}\rho) e^{ik_{||,mn}z} \\ \nonumber
& = &  u_{mn}(\vec r_{\perp}) \exp\left[{i \left(k-\frac{k_{\perp,mn}^{2}}{2k_{0}} \right)z}\right],
\end{eqnarray}
where we have factored out the axial position~$z$.
Here, $N_{mn} = 1/\sqrt{\pi R^{2}J_{m+1}^{2}(\lambda_{mn})}$ is the normalization factor,
$k_{\perp,mn} = \lambda_{mn}/R$, $k_{||,mn} = \sqrt{k^2 - k_{\perp,mn}^{2}}$,
$\lambda_{mn}$ is the $n$-th zero of the Bessel function of order $m$ of the first kind~$J_{m}$,
$k_{0} = \omega_{0}/c$ is the wave vector of the quantum light with central frequency $\omega_0$ and speed $c$.
Since we are working in the paraxial regime, where the perpendicular wave numbers are small, $k_{\perp,mn} \ll k$,
we make the approximation $k_{||,mn} - k \approx - k_{\perp,mn}/(2k) \approx - k_{\perp,mn}/(2k_0)$.
Here, we have also imposed the boundary condition $f_{k,mn}(\rho=R,\phi,z) = 0$
that all fields vanish at a virtual cut-off radius $R$. Thus we obtain a discrete set of transverse mode functions 
$u_{mn}(\vec r_{\perp})$ that is complete and orthogonal
\begin{eqnarray}
\int d^2\vec r_{\perp} u_{mn}^{*}(\vec r_{\perp}) u_{m'n'}(\vec r_{\perp}) = \delta_{mm'} \delta_{nn'},\\
\sum_{mn} u_{mn}^{*}(\vec r_{\perp}) u_{mn}(\vec r'_{\perp}) = \delta^{(2)}(\vec r_{\perp} - \vec r'_{\perp}).
\end{eqnarray}
In principle, one can use an arbitrary basis, but we choose this particular basis
where one can factorize the dependence on the longitudinal $z$ and transverse position $\vec r_\perp$, 
$f_{k,mn}(\vec r) = u_{mn}(\vec r_\perp) \exp \left[{i \left(k-\frac{k_{\perp,mn}^{2}}{2k_{0}} \right)z}\right]$.
As we will show below, this makes the numerical calculations of the problem manageable. 

The expression for the electric field of the quantum light $\vec E_{\text q}$ expanded on this basis set $f_{k,mn}(\vec r)$ is then given by
\begin{eqnarray}\nonumber
\vec E_{\text q}(\vec r) & = & \sqrt{\frac{2\pi \omega_0}{l}} \sum_{mn,k} \left[ \vec e_{\text q} f_{k,mn}(\vec r) \; a_{k,mn} + \text{H.c.}  \right],
\end{eqnarray}
where $l$ is the length of the quantization volume and $\vec e_{\text q}$ is the polarization of the quantum field.
We assume that both quantum and classical light fields are narrowband fields centered at $\omega_{0} = \omega_{e0} - \Delta$
and $\omega'_{0} = \omega_{e1} - \Delta$, respectively, so that $\omega_{k} \approx \omega_{0}$.
Here, $\Delta$ is the detuning from the excited state for both light fields with the corresponding atomic transition frequencies
$\omega_{e0}$ and $\omega_{e1}$.
We introduce slowly varying light operators for the quantum light field, which depend on time and position
\begin{eqnarray}
a_{mn}(z,t) & = & \sqrt{\frac{c}{l}} \sum_{k} a_{k,mn} e^{i \left( k - k_{0} -\frac{k_{\perp,mn}^{2}}{2 k_{0}} \right) z} e^{i\omega_{0} t}.
\end{eqnarray}
The annihilation operators of the light field $a_{mn}(z,t)$ are the expansion coefficients into 
the transverse mode basis and have the commutation relation
\begin{equation}
\left[ a_{mn}(z), a^{\dag}_{m'n'} (z') \right] = c \delta_{mm'} \delta_{nn'} \delta(z-z').
\end{equation}
With the chosen normalization, where $c$ appears in the commutator, the photon intensity in mode $mn$
at the position $z$ is given by $a^\dag_{mn}(z)a_{mn}(z)$.
The electric field rewritten by means of these new light operators,
\begin{eqnarray}\nonumber
\vec E_{\text q}(\vec r) & = & \sqrt{\frac{2\pi \omega_0}{c}} \sum_{mn} \left[ \vec e_{\text q} u_{mn}(\vec r_\perp) a_{mn}(z,t)
e^{ik_0 z - i\omega_0 t}+ \text{H.c.}  \right],
\end{eqnarray}
is now an expansion on the set of transverse mode functions $u_{mn}(\vec r_{\perp})$.

We also introduce slowly varying operators for the atomic quantities of interest, which depend on position and time:
\begin{eqnarray}
S(\vec r,t) & = & \sum_{i} \frac{1}{\sqrt{n(\vec r)}} \delta^{(3)} (\vec r-\vec r_{i}) \sigma_{10}^{(i)} \\ \nonumber
&& \times e^{-i(\omega_{0}-\omega'_{0})t +i(k_{0}-k'_{0})z}, \\
P(\vec r,t) & = & \sum_{i} \frac{1}{\sqrt{n(\vec r)}} \delta^{(3)} (\vec r-\vec r_{i}) \sigma_{e0}^{(i)} e^{-i\omega_{0} t +i k_{0} z}, \\
\sigma_{e1} (\vec r,t) & = & \sum_{i} \frac{1}{\sqrt{n(\vec r)}} \delta^{(3)} (\vec r-\vec r_{i}) \sigma_{e1}^{(i)} e^{-i\omega'_{0} t +i k'_{0} z}, \\
\sigma_{ee} (\vec r,t) & = & \sum_{i} \frac{1}{\sqrt{n(\vec r)}} \delta^{(3)} (\vec r-\vec r_{i}) \sigma_{ee}^{(i)}.
\end{eqnarray}
Here, $k'_{0} = \omega'_{0}/c$ is the wave number of the classical control field and $\sigma^{(i)}_{nn'} = |n\rangle_{(i)} \langle n'|$
are the internal state operators of the $i$-th atom.
$S(\vec r,t)$ denotes the collective spin-wave excitation in the atomic ensemble, $P(\vec r,t)$ is the polarization/coherence between 
the ground state $|0\rangle$ and the excited state $|e\rangle$, 
$\sigma_{e1}(\vec r,t)$ describes the coherence between the second ground state $|1\rangle$ and the excited state $|e\rangle$,
and $\sigma_{ee}(\vec r,t)$ is the occupation of the excited state. 
For~weak excitations below saturation, this excited state operator can be neglected. We also assume that the atoms are stationary. 
The normalization of the atomic operators is chosen so as to achieve the same-time commutation relations are
\begin{eqnarray}
\left[ S(\vec r,t), S^{\dag}(\vec{r'},t) \right] & = & \left[ P(\vec r,t), P^{\dag}(\vec{r'},t) \right] = \delta^{(3)}(\vec r - \vec{r'}),\\ 
\left[ P(\vec r,t), S^{\dag}(\vec{r'},t) \right] & = & \frac{1}{n(\vec r)} \sigma_{e1}(\vec r) \delta^{(3)}(\vec r - \vec{r'}).
\end{eqnarray}
It will be convenient to express the atomic operators in the same basis as the light operators. 
Therefore, we also expand the operators for the spin-wave and the polarization on the transverse mode basis
\begin{eqnarray}
S(\vec r,t) & = & \sum_{mn} u_{mn}^{*}(\vec r_{\perp}) S_{mn} (z,t),\\
P(\vec r,t) & = & \sum_{mn} u_{mn}^{*}(\vec r_{\perp}) P_{mn} (z,t).
\end{eqnarray}

Having defined and expanded the slowly varying atomic and light operators, we now consider the Hamiltonian for the system.
The free light Hamiltonian reads
\begin{equation}
H_{\rm L} = \sum_{k,mn} \omega_{k}^{\phantom{\dag}} a_{k,mn}^{\dag} a_{k,mn}^{\phantom{\dag}}
\end{equation}
with the linear dispersion relation $\omega_k = c k$. The atoms in the ensemble are described by the Hamiltonian
\begin{equation}\label{ham:a}
H_{\rm A} = \sum_{i} \Delta |e \rangle_{i} \langle e|,
\end{equation}
where the transformation to the rotating frame has been performed with respect to the laser frequency $\omega_0$.

In order to obtain the Hamiltonian describing the interaction of light with the entire atomic ensemble,
we start out by considering the coupling of light and a single atom in the dipole approximation
\begin{equation}\label{ham:la1}
H_{\rm L-A}^{(i)} = - \vec E_i \cdot \vec D_i,
\end{equation}
where $\vec E_i$ is the electric field and $\vec D_i$ denotes the electric dipole operator for the $i$-th atom.
The electric field consists of the quantum light field coupling states $|0\rangle$ and~$|e\rangle$, 
$E_{\mathrm{Q}}^{(\pm)} = E_{0\mathrm{Q}}^{(\pm)} e^{\mp i \omega_0 t}$,
and the classical field coupling states $|1\rangle$ and $|e\rangle$,
$E_{\mathrm{C}}^{(\pm)} = E_{0\mathrm{C}}^{(\pm)} e^{\mp i \omega'_0 t \pm i k_0' z}$.
The effective light-atom Hamiltonian for the atomic ensemble is obtained by summing all the single atom Hamiltonians in Eq.~(\ref{ham:la1}),
$H_{\rm L-A} = \sum_i H_{\rm L-A}^{(i)}$.
We introduce the spatially slowly varying operators and integrate over space whereby the sum over $i$ is absorbed in the operator definitions.
Within the rotating wave approximation, the final 
Hamiltonian describing the atomic ensemble and its coupling to the light fields then reads
\begin{widetext}
\begin{equation}\label{ham:la2}
H_{\rm L-A} = \Delta \int d^{3}\vec r \sqrt{n(\vec r)} \sigma_{ee}(\vec r, t) - \left[
\frac{1}{2}  \int d^{3}\vec r \; \Omega(\vec r,t) \sigma_{e1} (\vec r,t) -g_{0} \sum_{mm' nn'} B_{m'n',mn} \int d z P_{m'n'} (z,t) a_{mn} (z,t) + \rm{H. c.}\right],
\end{equation}
\end{widetext}
where $g_{0} = D_{0} \sqrt{2\pi \omega_{0} n_0/c}$ is the coupling constant for the quantum light field with 
$D_0 = \vec D_{e,0}^{(-)} \cdot \vec e_q$, and $\Omega(\vec r,t)$ is the slowly varying Rabi frequency
\begin{equation}
\Omega(\vec r,t) = 2 \vec D_{e,1}^{(-)} \cdot \langle \vec E_{{0{\rm C}}}^{(+)}(\vec r) \rangle .
\end{equation}
The first term of the Hamiltonian~(\ref{ham:la2}) is the atomic part and the second term describes the coupling to the strong classical control field.
The last term represents the interaction between the atomic ensemble and the quantum light.
We see from Eq.~(\ref{ham:la2}) that the collective enhancement leads to an increase of the coupling constant between atoms and light
by a factor of $\sqrt{n_{0}} \sim \sqrt{N_{A}}$ up to $g_{0} = g \sqrt{n_{0}}$, where $g$ is the coupling corresponding to a single atom.

The geometry of the ensemble will lead to coupling between different modes $mn$ and $m'n'$, 
which is described by the coupling matrix
\begin{equation}\label{eq:bb}
B_{mn,m'n'} = \frac{1}{n_0} \int d^{2} \vec r_{\perp} u_{mn}^{*} (\vec r_{\perp}) u_{m'n'} (\vec r_{\perp}) n(\vec r_\perp).
\end{equation}
In general, the coupling matrix $B_{mn,m'n'}$ will be $z$ dependent, but here we have avoided this by choosing 
a basis set $f_{k,mn}(\vec r)$ [Eq.~(\ref{eq:ff})]  in which one can factorize the dependence on the transverse $\vec r_\perp$ 
and the simple exponential dependence on the longitudinal position $z$.
Moreover, as mentioned earlier, we have chosen the atomic density distribution to be constant in the $z$ direction,
which leads to a $z$ independent coupling matrix $B_{mn,m'n'}$.
This significantly reduces the numerical complexity in the calculations below. 
In addition, the rotational symmetry implies that there is no coupling for different values of the azimuthal quantum number $m$, 
$B_{mn,m'n'} = B_{nn'}^{(m)} \delta_{mm'}$. Thus, we have an independent set of equations of motion for each $m$ value.
Note that in the one-dimensional limit, where $\sigma_\perp \rightarrow \infty$, 
the atomic density distribution is constant $n_{\rm 1D} = n_0$
and then the coupling matrix $B_{nn'}^{(m)}$ transforms to a unit matrix due to the orthogonality
of the $u_{mn} (\vec r_{\perp})$ modes, hence uncoupled modes. 

We assume for simplicity, that the classical driving field is spatially uniform with Rabi frequency $\Omega(\vec r,t) = \Omega(t)$.
It was shown in Ref.~[\onlinecite{Surmacz2008}], that the control field should be loosely focused with
beam waist at least more than twice that of the input light mode, which agrees with our assumption of the plane wave control field. 
Otherwise, the position dependent control field will lead to a position dependent AC Stark shift,
which may lead to reduced efficiency of the light storage. 

The above Hamiltonian~(\ref{ham:la2}) allows us to derive the equations of motion for
describing the propagation of light through the atomic ensemble as well as the evolution of the atomic coherences.
We use the Heisenberg equation and calculate the commutators to obtain
\begin{eqnarray}
\left( \frac{d}{dt} + c \frac{d}{dz} \right) a_{mn}(z,t) & = & -ic \frac{k_{\perp,mn}^{2}}{2k_{0}} a_{mn}(z,t) \\ \nonumber
&& + i g_{0} c \sum_{n'} B_{nn'}^{(m)} P_{mn'} (z,t), \\
\frac{d}{dt} P_{mn} (z,t) & = & -\left( \frac{1}{2} \gamma + i\Delta \right)  P_{mn} (z,t) \\ \nonumber
&& + \frac{1}{2} i \Omega(t) S_{mn} (z,t) \\ \nonumber
&& + i g_{0} \sum_{n'} B_{nn'}^{(m)} a_{mn'} (z,t), \\
\frac{d}{dt} S_{mn} (z,t) & = &  \frac{1}{2} i \Omega^{*}(t) P_{mn} (z,t).
\end{eqnarray}

The procedure employed here applies to a collection of near paraxial modes. 
In reality one should include all light modes. The non-paraxial modes would be needed
if one were to account for the evolution of all spontaneously emitted photons,
whereas here we will only be tracking the collectively enhanced emission.
It was shown in Ref.~[\onlinecite{Sorensen2008}] that for a dilute atomic ensemble in the ideal 
gas approximation the effect of non-paraxial modes can be accounted for by the inclusion of the polarization decay 
due to spontaneous emission at a rate~$\gamma$,
which has already been included in the expressions above. 
We assume that there is no spin-wave decay that leads to a loss of the spin-wave coherence $S (z,t)$.
It was shown in~Ref.~[\onlinecite{Gorshkov2007a}], that such a decay introduces a simple exponential decay
into the solutions for storage and read-out and does not make the optimization harder.
We also omit the quantum noise. In Ref.~[\onlinecite{Gorshkov2007a}], it was argued that
under reasonable experimental conditions, it is not necessary for calculating any normally ordered product such as the efficiency.
Alternatively, if required the noise can be reintroduced at a later stage. 
Thus the only two loss channels present in our model are the spontaneous decay $\gamma$ of the excited state 
in other directions than the collectively enhanced one
and the leakage of the input light mode out of the atomic ensemble. 
The latter one results from the fact that not all the light is stored within the atomic ensemble,
since part of the light will expand beyond the atomic ensemble $z>L$ during the storage process and hence be lost. 
For instance, avoiding this leakage in the resonant EIT case 
requires reduction of the group velocity $v_g \sim \Omega^2 L/(d \gamma)$ which, however,
increases the spontaneous decay. As a consequence the optimal memory performance
will be a trade-off between these two loss sources.

Next, we transform the equations of motion into a co-moving frame with $t' = t-z/c$
and introduce a dimensionless time $\tit = \gamma t'$, detuning $\tilde \Delta = \Delta /\gamma$, position $\zt = z/L$, 
and Rabi frequency $\tilde \Omega(\tit) = \Omega(\tit) /\gamma$. 
In addition, we introduce two dimensionless parameters describing the properties of the sample:
the peak optical depth
\begin{equation}
d_0 = \frac{4L|g_{0}|^{2}}{\gamma}
\end{equation}
and the Fresnel number of the atomic ensemble
\begin{equation}
F = \frac{\sigma_\perp^2}{L \lambda_0},
\end{equation}
where $\lambda_0$ is the wavelength of quantum light. 
The optical depth quantifies the absorption of resonant light $e^{-d}$ and thereby describes the strength of the coupling between
the light and atomic ensemble. The Fresnel number characterizes the geometry of the atomic ensemble with regard to the diffraction of light. 
With this we obtain the dimensionless coupled equations describing the atom-light system
\begin{eqnarray}\label{eq:a}
\frac{d}{d\zt} a_{mn}(\zt,\tit) & = & -i \frac{k_{\perp,mn}^2 \sigma_\perp^2}{4 \pi F} a_{mn}(\zt,\tit) \\ \nonumber
&& + \frac{1}{2} i \sqrt{d_{0}}  \sum_{n'} B_{nn'}^{(m)} P_{mn'} (\zt,\tit), \\ \nonumber
\frac{d}{d\tit} P_{mn} (\zt,\tit) & = & -\left( \frac{1}{2} + i\tilde \Delta \right)  P_{mn} (\zt,\tit) + \frac{1}{2} i \tilde\Omega(\tit) S_{mn} (\zt,\tit) \\ \label{eq:p}
&& + \frac{1}{2} i \sqrt{d_{0}} \sum_{n'} B_{nn'}^{(m)} a_{mn'} (\zt,\tit), \\ \label{eq:s}
\frac{d}{d\tit} S_{mn} (\zt,\tit) & = &  \frac{1}{2} i \tilde \Omega^{*}(\tit) P_{mn} (\zt,\tit).
\end{eqnarray}
Here, a factor of $1/\sqrt{\gamma}$ has been absorbed into $a_{mn}(\zt,\tit)$ and a factor of $\sqrt{L}$ into the definitions
of $P_{mn} (\zt,\tit)$ and $S_{mn} (\zt,\tit)$.
One can see from the above equations of motion that the only two physical parameters that describe the quantum interface
between light and matter are the peak optical depth~$d_0$ and the Fresnel number~$F$.
Since the perpendicular wave number scales as  $k_{\perp,mn} \sim 1/\sigma_\perp$
so that the first term in Eq.~(\ref{eq:a}) only depends on the Fresnel number~$F$.
It will be shown in Section~\ref{sec:read} that the parameters of the control field 
(its temporal shape $\tilde \Omega(t)$ and detuning~$\tilde \Delta$) do not influence the efficiencies of the light retrieval.

In our model, we neglect the fact that the spontaneously emitted photons can be rescattered
and interfere with the quantum light signal. 
However, this effect is negligible in the paraxial regime, where the solid angle of emitted light is small.
Moreover, the storage we are interested in employs only a single or a few photon pulses,
thus in this case there will be at most a few spontaneously emitted photons.
In addition, the probability of the spontaneous emission process is small in most experiments and decreases with increasing optical depth.

Before proceeding, we can relate the three-dimensional theory to the one-dimensional one \cite{Gorshkov2007a},
where one assumes that the atomic density distribution is independent of the transverse coordinate $n(\vec r) = n(z)$
[for our choice of the atomic density it corresponds to $n(\vec r) = n_0$],
which leads to an infinite value of the Fresnel number of the atomic ensemble $F\rightarrow \infty$.
The first term in the propagation equation~(\ref{eq:a}) gives either a phase depending on the momentum
of the incoming quantum light signal or it vanishes if the waist of the incoming beam is wide enough. 
In the 1D limit, the coupling matrix $B_{nn'}^{(m)}$ in Eq.~(\ref{eq:bb}) becomes the identity matrix, 
since the modes $u_{mn}(\vec r_\perp)$ are orthogonal.
In consequence, the last terms in Eqs.~(\ref{eq:a}) and~(\ref{eq:p}) transform into $\frac{1}{2}i\sqrt{d_0}P_{mn}(\zt,\tit)$
and $\frac{1}{2}i\sqrt{d_0}a_{mn}(\zt,\tit)$, respectively. Since there is no longer any coupling between distinct transverse modes,
we treat each mode independently. 
In this manner, we end up with the one-dimensional equations from Ref.~[\onlinecite{Gorshkov2007a}], 
where the only important parameter is the optical depth $d_0$. The typically employed one-dimensional description thus  applies 
to all transverse modes in the limit of $F\rightarrow\infty$.

%%%%%%%%%%%%%%%%%%%%%%%%%%%%%%%%%%%%%%%%%%
\section{Optimal read-out from the memory}\label{sec:read}
%%%%%%%%%%%%%%%%%%%%%%%%%%%%%%%%%%%%%%%%%%

In this section, we study the optimal retrieval of light stored in a quantum memory based on a $\Lambda$-type atomic ensemble.
The separate analysis of the read-out process is not only interesting for the quantum memory purposes
but also for gaining knowledge about quantum states of atoms generated by other means.
Such atomic states can be mapped onto light states in order to study their properties or quantum correlations~\cite{Pohl2010}.
As noted above, the equations of motion describing the quantum light-matter interface [Eqs.~(\ref{eq:a})-(\ref{eq:s})]
depend only on two physical parameters of the atomic ensemble, $F$ and $d_0$. 
In addition, the equations of motion also depend on the coupling to the control light field $\Omega(t)$ and its detuning $\Delta$ 
from the transition energy. In the one-dimensional theory, the retrieval efficiency is independent of the detuning 
and temporal shape of the control field.
In this section, we will first show that this holds also in the case of many coupled transverse modes.
Next, we study the dependence of the read-out efficiency on the Fresnel number of the atomic ensemble~$F$.

Initially all atoms are in the ground state and no atomic excitations are present,
which corresponds to an empty quantum memory. 
The photon storage is based on the mapping of the incoming quantum light field onto some spin-wave mode in the atomic ensemble.
At a later time, the stationary excitation is restored onto an outgoing light mode. 
The retrieval efficiency is then the ratio between the number of retrieved photons and the number of stored excitations.
If one normalizes the stored spin-wave according to $\sum_{mn} \int_0^1 d\zt |S_{mn}(\zt,\tit=0)|^2 = 1$,
the read-out efficiency is
\begin{equation}
\eta_{\rm r} = \sum_{mn} \int_{0}^{\infty} d\tit | a_{mn}(\zt = 1,\tit) |^{2} = \sum_{mn} \int_{0}^{\infty} d\tit | a_{mn}^{\mathrm{out}}(\tit) |^{2},
\end{equation}
where we have assumed that the retrieval starts at time $\tit=0$ 
and continues until all excitations are read out from the memory at $\tit=\infty$.

We want to prove that the read-out efficiency $\eta_{\rm r}$ depends neither on the detuning $\tilde\Delta$ 
nor on the temporal shape of the control field $\tilde \Omega(\tit)$. To this end we use the conservation law
\begin{equation}\label{eq:cons}
\frac{d}{d\zt} |\vec a(\zt,\tit)|^2 + \frac{d}{d\tit} |\vec P(\zt,\tit)|^2 + \frac{d}{d\tit} |\vec S(\zt,\tit)|^2 = - |\vec P(\zt,\tit)|^2,
\end{equation}
which states that the number of excitations in the system can change only due to the spontaneous emission, 
since this is the only decoherence channel for the read-out process. This can be derived directly from Eqs.~(\ref{eq:a})-(\ref{eq:s}).
Here, we have for simplicity introduced a compact vector notation $\sum_{n} a_{mn}(\zt,\tit) \rightarrow \vec a^{(m)}(\zt,\tit)$,
which replaces the mode notation with indices and sums.
Next, we integrate Eq.~(\ref{eq:cons}) over space and time $\left (\int_0^1 d\zt \int_0^\infty d\tit\right)$.
We use the initial conditions that there is a normalized spin-wave excitation in the ensemble $\vec S(\zt,\tit=0) = \vec S_0(\zt)$, 
$\int_0^1 d\zt |\vec S_0(\zt)|^2=1$ and no coherence between the ground and excited states $\vec P(\zt,0)=0$,
as well as the boundary condition that there is no incoming quantum light field during the read-out procedure: $\vec a_{\rm in}(\zt =0,\tit)= 0$.
We also assume that we deal with a complete retrieval, where no excitations are left in the system after the read-out is finished:
$\vec S(\zt,\infty) = \vec P(\zt,\infty) = 0$, which can be ensured by sending in sufficiently strong control light. 
Then we find that the retrieval efficiency also can be expressed as unity minus the losses during the procedure
\begin{eqnarray}\label{eq:ret}
\eta_{\rm r} & = & 1 - \int_{0}^{1} d \zt \int_{0}^{\infty} d\tit \; | \vec P(\zt,\tit) |^{2} \\ \nonumber
& = & 1 - \int_0^1d\zt \; \vec l(\zt),
\end{eqnarray}
where $\vec l(\zt)$ is a position-dependent loss per unit length:
\begin{eqnarray}
\vec l(\zt) & = & \int_0^{\infty} d\tit \; |\vec P(\zt,\tit) |^{2} \\ \nonumber
&&  = \mathcal{L}^{-1} \left[ \int_0^{\infty} d\tit \vec P^\dag(u',\tit) \vec P(u,\tit) \right],
\end{eqnarray}
and $\mathcal{L}^{-1}$ denotes double inverse Laplace transform ($u \rightarrow~\zt, u' \rightarrow \zt'$) 
in the position coordinate, $f(u) = \mathcal{L}[f(\zt)] = \int_0^1 d\zt e^{-u\zt} f(\zt)$, evaluated at position $\zt$ in both cases.
In the following part of the paper we will use the purely imaginary $u$ and transform to $u \rightarrow iu$, 
where from now on $u$ is purely real. 
Next, we wish to express the loss in terms of the Laplace transformed spin-wave $\vec S_0(u) = \vec S(u,\tit=0)$ by a relation of the form
\begin{equation}
 \vec l (\zt) = \mathcal{L}^{-1} \left[ \vec S_0^\dag(u') \mathbb{A}(u',u) \vec S (u) \right].
 \end{equation}
In view of the boundary conditions, this can be achieved if we can find a matrix $  \mathbb{A}(u,u')$ that fulfills
\begin{eqnarray} \label{eq:ppp}\nonumber
 \frac{d}{d\tit} \left[ \vec P^\dag(u',\tit)  \mathbb{A}(u',u) \vec P(u,\tit) + \vec S^\dag(u',\tit)  \mathbb{A}(u',u) \vec S(u,\tit)  \right]  \\ 
 = \vec P^\dag(u',\tit) \vec P(u,\tit). \phantom{adsfasdfasf}
\end{eqnarray}
The compact notation of the matrix $\mathbb{A}(u',u)$ is shorthand for a four-dimensional matrix with components $A_{nn'}^{(m)}(u',u)$.
In Appendix~\ref{app:proof}, we prove that it is indeed always possible to find a matrix $\mathbb{A}(u',u)$ fulfilling Eq.~(\ref{eq:ppp}).
This allows us to rewrite the position dependent loss in terms of the spin-wave $\vec S_0(u)$ and the matrix $\mathbb{A}(u,u')$,
which only depends on the optical depth $d_0$ and the Fresnel number $F$ while independent of the detuning $\tilde \Delta$
and the temporal shape of the driving field $\tilde \Omega(\tit)$. The loss is proportional to the ratio between the
decay of the spin-wave into the unwanted directions and the constructively interfering decay into the desired output light pulse
with a direction set by that of the control beam.
Note that in the case of nonzero spin dephasing $\gamma_s$, which introduces exponential decay of the spin-wave
and the corresponding output light, the loss becomes dependent on the control parameters 
and the shape of the output pulse~\cite{Gorshkov2007a}.

\begin{figure}[t] 
\begin{center} 
\unitlength 1mm
{\resizebox{87mm}{!}{\includegraphics{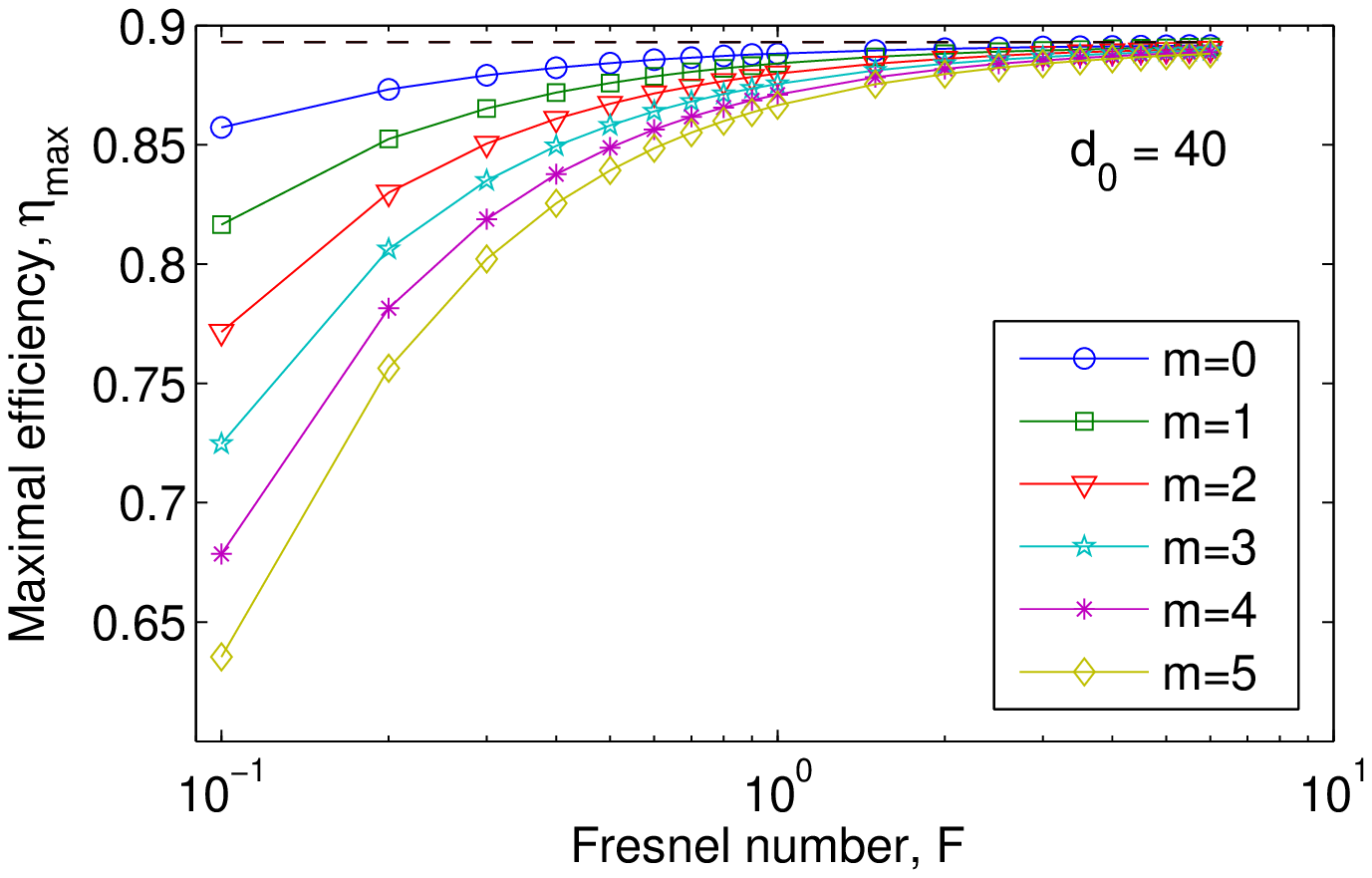}}}
{\resizebox{87mm}{!}{\includegraphics{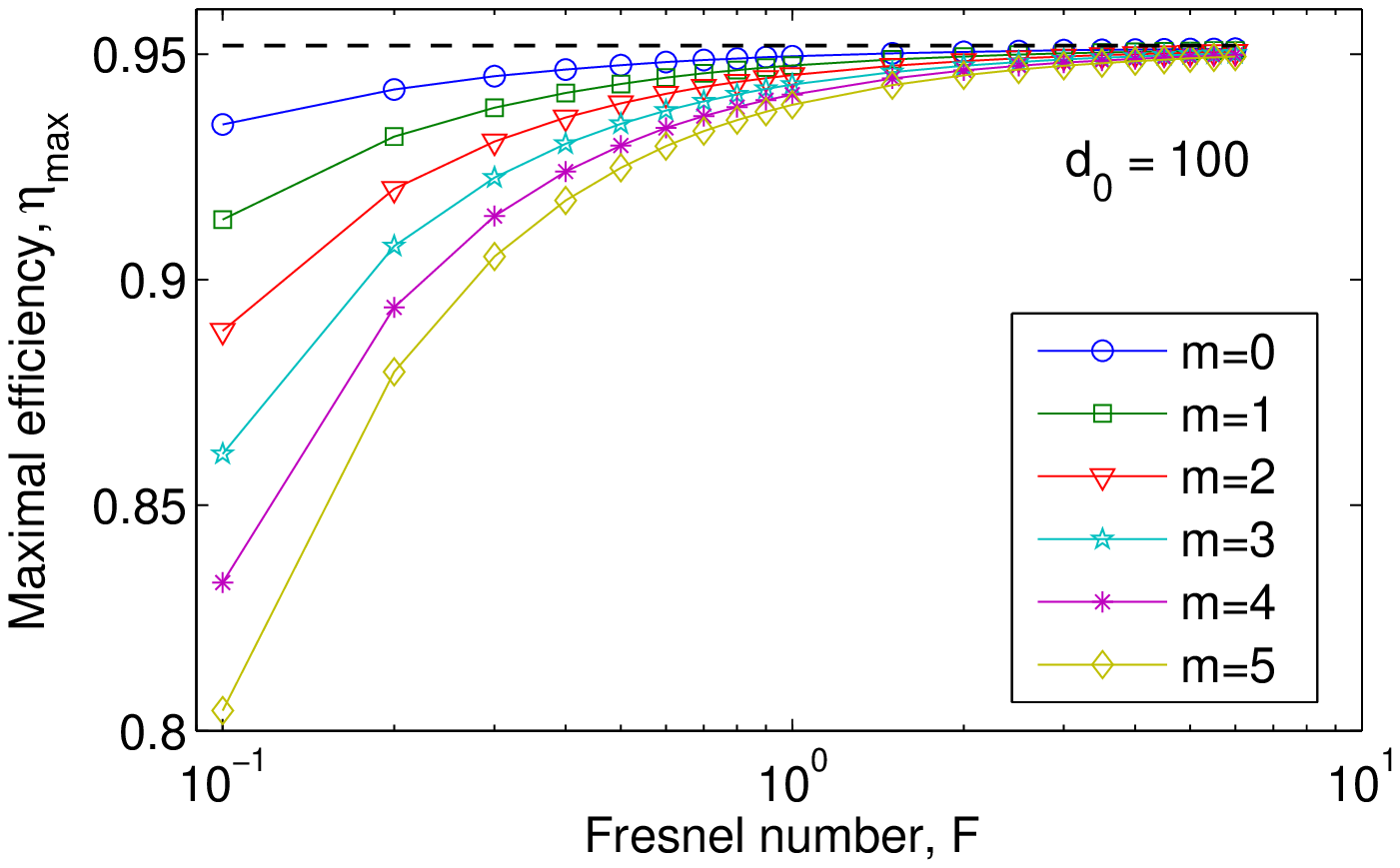}}}
\end{center} 
\caption{\label{fig:max}(Color online) Maximum retrieval efficiency as a function of the Fresnel number of the atomic ensemble~$F$, 
for different transverse azimuthal quantum numbers $m$, and for fixed peak optical depth $d_0 = 40$ (top) 
and $d_0 = 100$ (bottom). The dashed black lines denote the 1D limit of $F\rightarrow \infty$.}
\end{figure}

After finding the matrix $\mathbb{A}(u,u')$ as the solution of the Sylvester equation~(\ref{eq:syl}), we can rewrite the expression 
for the read-out efficiency~(\ref{eq:ret}) as a matrix product corresponding to an eigenvalue problem
\begin{eqnarray}\label{eq:eig}
\eta_\mathrm{r} & = & \int_0^1 d\zt \mathcal{L}^{-1} \left\{ \vec S_0^\dag(u') \left[ \mathbb{I}-\mathbb{A}(u',u) \right] \vec S_0 (u) \right\}\\ \nonumber
& = &  \int_{-\infty}^{\infty} du \int_{-\infty}^{\infty} du' \vec S_0^\dag (u')  \left[ \frac{e^{i(u-u')}-1}{4\pi^2 i(u-u')} \right] \\ \nonumber
&& \times  \left[ \mathbb{I}-\mathbb{A}(u',u) \right]  \vec S_0(u).
\end{eqnarray}

In order to calculate the efficiency numerically, we discretize the Laplace transformed position $\zt \rightarrow u$ 
with a finite cut-off $u_{\rm max}$. The value of this cut-off and how fine the grid resolution needs to be depend both on the
physical parameters $d_0$ and $F$. 
A finite number of Bessel modes depending on the Fresnel number $F$ has been used
in order to set the coupling matrix $\mathbb{B}$ in Eq.~(\ref{eq:bb}) in the Sylvester equation~(\ref{eq:syl}).
This equation is solved numerically to find the matrix $\mathbb{A}(u',u)$.
The integral over the position $\zt$ can be calculated analytically. 
Next, we construct the matrix resulting from the inverse Laplace transform $[e^{i(u-u')}-1]/[4\pi^2 i(u-u')]$.
The two integrals over $u$ and $u'$ are replaced by sums and we re-index the four-dimensional matrix $\mathbb{A}$ 
so as to turn each pair of indices into one index $A_{nn'} (u,u') \rightarrow A_{\tilde n,\tilde n'}$, where $\tilde n = (n,u)$ and $\tilde n' = (n',u')$.
Thus, the sums of the four-dimensional matrices can be implemented as multiplications of two-dimensional matrices. 
The retrieval efficiency is thus an eigenvalue of the large kernel matrix 
$\mathbb{C}_r \sim \left[\mathbb{I}-\mathbb{A}(u',u) \right] [e^{i(u-u')}-1]/[4\pi^2 i(u-u')]$ with the corresponding eigenstate $\vec S_0(u)$. 
The optimized spin-wave to retrieve from is the eigenvector corresponding to the highest eigenvalue.
In turn, we can determine the corresponding optimal outcoming light mode.
An alternative method of calculating the retrieval efficiency (described in detail in the next section) 
involves the adiabatic elimination of the excited state followed by calculations with the Laplace transform in the time variable~$\tit$. 
It is worth mentioning that this optimal spin-wave to read-out from
is not necessarily the best one for the combined process of write-in and read-out.
Thus the full memory procedure has to optimized, which is done in detail in Sec.~\ref{sec:full}.

In the case of the one-dimensional theory, the read-out efficiency depends only on one physical parameter: the optical depth $d_0$ as 
$\eta_\mathrm{r}  \sim 1 - 1/d_0$ in the $d_0 \rightarrow \infty$ limit. 
Here, we include an additional degree of freedom that describes the geometry of the atomic ensemble 
with regards to the diffraction of light, namely the Fresnel number of the atomic ensemble~$F$.

We have optimized the read-out process from the memory numerically for two fixed values of the peak optical depth $d_0 = 40$ and $d_0=100$
and we have analyzed the influence of the Fresnel number~$F$.
Here, the memory operates on resonance, $\tilde \Delta = 0$.
The maximum retrieval efficiency for different branches of the azimuthal quantum number $m$ is shown in Fig.~\ref{fig:max}
for $d_0=40$ (top) and $d_0 = 100$ (bottom).
The results are plotted together with the one-dimensional case of infinitively large Fresnel number, $F \rightarrow \infty$ (dashed black lines).
The maximal retrieval efficiency of the spatial memory approaches this limit already for small $F \sim 1$ for several values of~$m$.
The figure clearly shows the spatial multimode character of the memory, since high efficiencies for different values of the 
azimuthal quantum number $m$ exist. Furthermore, in the plot we show only the optimal mode for each~$m$.
In general, there are several orthogonal modes with high efficiencies within each $m$-subspace.

\begin{figure}[t] 
\begin{center} 
\unitlength 1mm
{\resizebox{87mm}{!}{\includegraphics{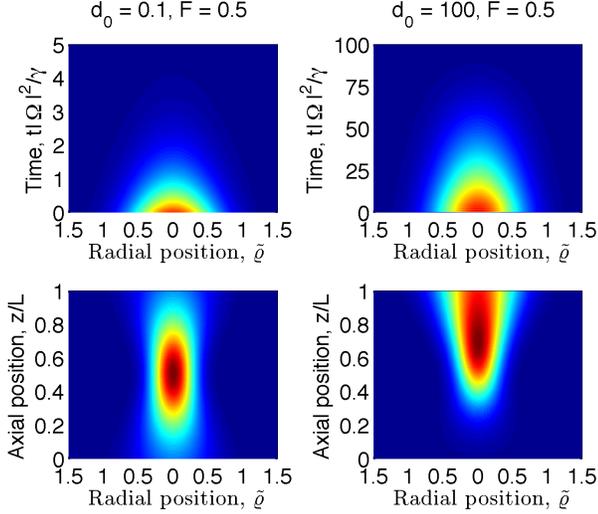}}}
\end{center} 
\caption{\label{fig:swa}(Color online) (top) The intensity of the optimal output light modes for retrieval at the end of the atomic ensemble, $\zt=1$
for (left) low $d_0=0.1$ and (right) high $d_0=100$ peak optical depths.
(bottom) The corresponding density of the optimal spin-wave excitations stored in the atomic memory. 
In both cases, the Fresnel number of the ensemble is $F=0.5$ and retrieval is resonant $\tilde \Delta=0$.
The radial position is scaled with the atomic ensemble width as $\tilde \rho = \rho/\sigma_\perp$ and time according to
$\tit = t |\Omega|^2/ \gamma$.}
\end{figure}

In Fig.~\ref{fig:swa}, we show the density of the optimal spin-wave excitation to read-out from
$|\vec S_{\rm opt}(\tilde \rho,\zt)|^2$ and the corresponding optimal output light intensity 
$|\vec a_{\rm out}^{\rm \; opt}(\tilde \rho,\tit)|^2$, where $\tilde \rho = \rho/\sigma_\perp$.
Here, we show the results for the Fresnel number $F=0.5$ and for low ($d_0=0.1$) and high ($d_0=100$) values of the peak optical depth. 
For simplicity, we assumed a constant Rabi frequency of the classical light field $\tilde \Omega$.
Note that the radial position is scaled with respect to the width of the atomic ensemble $\tilde\rho = \rho/\sigma_\perp$
and time is dimensionless and scaled according to $\tit = t |\Omega|^2/ \gamma$. 
In the case of a low absorbing medium, $d_0=0.1$ [Fig.~\ref{fig:swa}, (left)], 
the spin-wave is spread over the entire ensemble length and its density is symmetric with respect to the middle 
of the sample in the axial position~$\zt$.
For high optical depth $d_0=100$ [Fig.~\ref{fig:swa}, (right)], most of the excitation is localized closer the end of the sample ($\zt=1$)
from which the light is read out. 
In the latter case, the excitation does not need to propagate through the entire length of the sample, which in consequence
leads to lower losses due to spontaneous emission. 
Thus, if the optical depth is high enough it is better to effectively use only a part of the ensemble
and thereby decrease the propagation length.
The distribution of the optimal spin-wave excitation in the transverse direction reflects the atomic distribution,
but also depends on the Fresnel number of the ensemble,
which affects the diffraction of the input and output light pulses (see detailed discussion in Section~\ref{sec:results}).

The transverse shape of the corresponding optimal output light pulse also reflects the Gaussian distribution of the atoms.
However, its duration and temporal shape strongly depend on the peak optical depth~$d_0$, which was studied in detail
in the 1D theory~\cite{Gorshkov2007a}. For the resonant case under consideration, $\tilde \Delta=0$, the output pulse length is proportional to 
the optical depth $d_0$, since it is proportional to the inverse of the group velocity $t_{\rm out} \sim 1/v_g \sim d_0\gamma/|\Omega|^2$.
Note that for the off-resonant Raman transition ($\Delta \gg \gamma d_0$), 
the pulse duration dependence on the optical depth is different,
$t_{\rm out} \sim \Delta^2/(d_0\gamma|\Omega|^2)$.
Since the retrieval efficiency is detuning independent, the choice of the experimental realization may be affected
by the available laser power and the length of the pulses one wants to retrieve.
Unlike the efficiency, the shape of the light pulse does depend on the detuning $\tilde \Delta$.

%%%%%%%%%%%%%%%%%%%%%%%%%%%%%%%%%%%%%%%%%%
\section{Adiabatic storage and retrieval}\label{sec:full}
%%%%%%%%%%%%%%%%%%%%%%%%%%%%%%%%%%%%%%%%%%

\begin{figure}[tb]
\begin{center} 
\unitlength 1mm
{\resizebox{89mm}{!}{\includegraphics{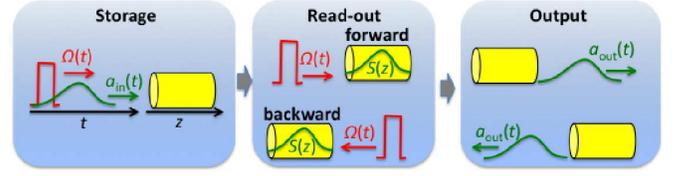}}}
\end{center} 
\caption{\label{fig:qm_scheme}
(Color online) Schematic illustration of the two operation modes of a quantum memory: (top) forward, 
where the control fields $\Omega(t)$ during the storage and read-out
are in the same direction, and (bottom) backward read-out with the opposite directions of the control fields.}
\end{figure}

In section \ref{sec:read}, we solved the full three-dimensional problem
for the read-out of a spin-wave from an atomic ensemble and studied its dependence on the crucial physical parameters:
the peak optical depth $d_0$ and the Fresnel number $F$.
In order to understand the operation of a full quantum memory, 
one also needs to store the light beforehand and we now turn to the full process of storage followed by retrieval. 
A quantum memory is often discussed in two different modes of operation (see Fig~\ref{fig:qm_scheme}):
forward operation, where the stored light is retrieved in the same direction as it was written in
and backward operation, where the read-out is performed in the opposite direction~\cite{Gorshkov2007a,Hammerer2010}.

%%%%%%%%%%%%%%%%%%%%%%%%%%%%%%%%%%%%%%%%%%
\subsection{Read-out in the forward direction}\label{sec:forw}
%%%%%%%%%%%%%%%%%%%%%%%%%%%%%%%%%%%%%%%%%%

In this subsection, we derive analytical expressions for the efficiency of storage 
followed by retrieval in the forward direction together with the input-output relations between 
the incoming and outcoming light modes and the spin-wave in the atomic ensemble.

For simplicity, we restrict ourself to the limit where we can adiabatically eliminate the excited state by setting
$\frac{d}{d\tit} \vec P(\zt,\tit)\approx 0$.
The condition for this is $t_{\rm pulse} d_0 \gamma \gg 1$, 
where $t_{\rm pulse}$ is the duration of the preferably smooth quantum light pulse.
The adiabatic elimination leads to the following equations of motion
(now written in compact vector/matrix notation)
\begin{eqnarray}\label{eq:adp}
\vec P(\zt,\tit) & = & -\frac{\frac{1}{2}i\tilde \Omega(\tit)}{\frac{1}{2}+i\tilde \Delta} \vec S(\zt,\tit)  
-\frac{\frac{1}{2}i\sqrt{d_0}}{\frac{1}{2}+i\tilde \Delta} \mathbb{B}
  \vec a(\zt,\tit), \\ \nonumber
  \frac{d}{d\zt} \vec a(\zt,\tit) & = & \left(-\frac{i \vec k_\perp^2 \sigma_\perp^2}{4\pi F} - \frac{\frac{1}{4}d_0}{\frac{1}{2}
+i\tilde \Delta} \mathbb{B}^2 \right)\vec a(\zt,\tit) \\ \label{eq:aa}
&& -\frac{\frac{1}{4}\sqrt{d_0}\tilde \Omega(\tit)}{\frac{1}{2}+i\tilde \Delta}\mathbb{B} \vec S(\zt,\tit), \\ \label{eq:pa}
\frac{d}{d\tit} \vec S(\zt,\tit) & = & -\frac{\frac{1}{4}|\tilde \Omega(\tit)|^2}{\frac{1}{2}+i\tilde \Delta} \vec S(\zt,\tit) 
-\frac{\frac{1}{4}\sqrt{d_0}\tilde \Omega^*(\tit)}{\frac{1}{2}+i\tilde \Delta} \mathbb{B} \vec a(\zt,\tit).
\end{eqnarray}
Here, the imaginary part of the first term in Eq.~(\ref{eq:aa}) is the phase shift due to the index of refraction of the medium with 
$n_{\rm refr} = -i \vec k_\perp^2 \sigma_\perp^2/(4\pi F) + d_0\tilde\Delta \mathbb{B}^2/(1+4\tilde \Delta^2)$.
Note that $\vec k_\perp^2$ although written as a vector is a diagonal matrix with $k_{\perp,mn}^2$ values on the diagonal. 
The AC Stark shift of the atoms is accounted for by the imaginary part of the first term in Eq.~(\ref{eq:pa}),
$\tilde\Delta|\tilde\Omega(t)|^2/(1+4\tilde \Delta^2)$ and is zero in the resonant case. 
The real parts of these two terms describe the damping of the light and spin-wave excitations
with the respective absorption coefficient $\frac{1}{2}d_0\mathbb{B}^2/(1+4\tilde\Delta^2)$
and effective decay rate $\frac{1}{2}|\tilde\Omega(t)|^2/(1+4\tilde\Delta^2)$   
due to the spontaneous emission at a rate~$\gamma$.
The last terms in Eqs.~(\ref{eq:aa}) and~(\ref{eq:pa}) represent the coherent interaction between light and the atomic ensemble.

It is worth noting that the control field $\tilde \Omega(\tit)$ determines the speed of the memory and can be completely eliminated
from the equations of motion by the rescaling: $\vec P(\zt,\tit) \rightarrow \vec P(\zt,\tit)/\tilde \Omega(\tit)$ 
and $\vec a(\zt,\tit) \rightarrow \vec a(\zt,\tit)/\tilde \Omega(\tit)$ and changing variables 
$\tit  \rightarrow v = h(0,\tit) = \int_0^{\tit} d\tit' |\tilde \Omega(\tit')|^2$,
where $h(0,\tit)$ is the integrated intensity of the control classical field. 
Thus, the dynamics of the system is independent of the temporal shape of the driving light field described by $\tilde\Omega(\tit)$,
which merely reshapes the shape of the incoming light mode which is stored and accordingly the shape of the output mode. 
However, here we do not perform the rescaling in order to avoid introducing additional notation. 

From the equations of motion (\ref{eq:aa}) and (\ref{eq:pa}), one can derive expressions
for the spin-wave excitation stored in the sample and for the output light for given incoming quantum light pulse.
We have solved these equations in two different ways. 
In the first case, we use the Laplace transform of the position $\zt$ and we obtain equations in momentum space $u$ and time $\tit$.
This transformation was also used in the previous section for proving that the read-out efficiency only depends 
on the optical depth $d_0$ and the Fresnel number $F$. 
The second approach is based on Laplace transform in time and leads to solutions in position space and frequency.
The numerical calculations were performed with both approaches yielding the same results.
Here, we only discuss in detail the analytical expressions for the forward read-out with Laplace transform in space. 
The alternative approach as well as the backward mode of operation are described in detail 
in Appendicies~\ref{app:for} and~\ref{app:back}, respectively. 

Solving the coupled equations of motion (\ref{eq:adp}), (\ref{eq:aa}), and (\ref{eq:pa}),
the connection between outcoming light, spin-wave, and incoming light can be written in the form of input-output beam splitter relations.
In addition to the initial and boundary conditions introduced for the read-out in Sec.~\ref{sec:read},
we use here the appropriate conditions for the storage procedure. 
There is no spin-wave excitations in the ensemble at the start of the writing process
$\vec S(\zt,\tit=-\infty) = 0$ and no coherence between the ground and excited state $\vec P(\zt,\tit=-\infty) =0$.
The storage of the incoming light field $\vec a(\zt = 0,\tit) = \vec a_{\rm in}(\tit)$ takes place in the time interval $\tit \in (-\infty,0]$
and is normalized according to $\int_{-\infty}^{0} d\tit |\vec a_{\rm in}(\tit)|^{2} = 1$.
We perform Laplace transform in space $\zt$, which leads to elimination of the differential equation~(\ref{eq:aa})
for the quantum light $\vec a(\zt,\tit)$. 
Thus, the above mentioned relations in momentum and time space read
\begin{eqnarray}
\vec a_{\mathrm{out}}(\tit) & = & \frac{1}{2\pi i} \int_{-\infty}^\infty du \; \mathbb{M}[\tilde \Omega(\tit),\tit,u] e^{iu} \; \vec S_0(u), \\ \label{eq:insw}
\vec S_0(u) & = & \int_{-\infty}^0 d\tit \; \mathbb{M}^T[\tilde \Omega^*(-\tit),-\tit,u] \; \vec a_{\mathrm{in}} (\tit).
\end{eqnarray}
Here, $\mathbb{M}[\tilde \Omega(\tit),\tit,u] = \mathbb{Q}(u) e^{ \mathbb{N}(u) \tit} $ with 
\begin{eqnarray}
\mathbb{Q}(u) & = & -\frac{\frac{1}{4}\sqrt{d_0}\tilde \Omega(\tit)}{\frac{1}{2}+i\tilde \Delta} \; \mathbb{T}^{-1}(u) \; \mathbb{B}, \\ \nonumber
\mathbb{N}(u) & = & \frac{-\frac{1}{4}|\tilde \Omega(\tit)|^2}{\frac{1}{2}+i\tilde \Delta} 
+ \frac{\frac{1}{16}d_0 |\tilde \Omega(\tit)|^2}{(\frac{1}{2}+i\tilde \Delta)^2} 
\mathbb{B} \; \mathbb{T}^{-1}(u) \; \mathbb{B}, \\
\mathbb{T}(u) & = & \left[ i u + i \frac{\vec k_{\perp}^2 \sigma_\perp^2}{4\pi F} + \frac{\frac{1}{4}d_0}{\frac{1}{2}+i\tilde \Delta} \mathbb{B}^2 \right],
\end{eqnarray}
and the coupling matrix $\mathbb{B}$ is defined in Eq.~(\ref{eq:bb}).

We start by considering the process of writing information into the atomic ensemble. 
The storage efficiency is the ratio between the number of stored excitations and the number of incoming photons. 
If the incoming light is normalized to one, the storage efficiency is given by
\begin{eqnarray}
\eta_{\rm s} & = & \int_0^1 d \zt |\vec S_0(\zt)|^2\\ \nonumber
& = &  \int_{-\infty}^{\infty} du \int_{-\infty}^{\infty} du' \left[ \frac{e^{i(u'-u)}-1}{4\pi^2 i(u'-u)} \right] \vec S_0^\dag (u) \vec S_0(u').
\end{eqnarray}
We substitute the corresponding input-output relation from Eq.~(\ref{eq:insw}) and obtain
\begin{eqnarray}\label{eq:etas}
\eta_s & = & \int_{-\infty}^\infty du \int_{-\infty}^\infty du'  \int_{-\infty}^0 d\tit  \int_{-\infty}^0 d\tit' \\ \nonumber
&& \times \vec a_{\mathrm{in}}^\dag(\tit)	 \left[ \frac{e^{i(u'-u)}-1}{4\pi^2 i(u'-u)} \right]  \mathbb{M}^*[\tilde \Omega^*(-\tit),-\tit,u]	\\ \nonumber
&& \times  \mathbb{M}^T[\tilde \Omega^*(-\tit'),-\tit',u'] \vec a_{\mathrm{in}} (\tit').
\end{eqnarray}

In order to numerically calculate the efficiencies, we discretize time $\tit$ and momentum $u$
and introduce finite cut-offs in both of them. The requirements for the cut-off values depend on the chosen peak optical depth
$d_0$ and the Fresnel number $F$.
Next, we construct the $\mathbb{M}$ matrix and the term resulting from the inverse Laplace transform.
The sizes of these depend on the number of discrete $\tit$ and $u$ points we take. 
Now, the integrals can be replaced by sums and, in consequence, can be implemented as matrix multiplications. 
In the end, we have a large kernel matrix
$\mathbb{C}_s [\tit,\tit'] \sim \mathbb{M}^*[\tilde \Omega^*(-\tit),-\tit,u] \mathbb{M}^T[\tilde \Omega^*(-\tit'),-\tit',u'] [ e^{i(u'-u)}-1]/[4\pi^2 i(u'-u)] $,
which after reindexation can be diagonalized in order to optimize the storage.
The eigenvalues of this matrix give the storage efficiencies with corresponding
incoming quantum light modes~$\vec a_{\mathrm{in}} (\tit)$. 
The highest eigenvalue is the maximal storage efficiency and the corresponding eigenvector is the optimized incoming light mode for storage.

Now, we turn to the full combined procedure of storage of an incoming light pulse followed by forward retrieval of the spin-wave.
The relation between the incoming and outcoming light at $\zt = 1$ is
\begin{eqnarray}\label{eq:ag}
\vec a_{\mathrm{out}}(\tit) & = & \frac{1}{2\pi i} \int_{-\infty}^\infty du  \int_{-\infty}^0 d\tit' \; e^{iu} \; \mathbb{M}[\tilde \Omega(\tit),\tit,u] \\ \nonumber
&& \times \mathbb{M}^T[\tilde \Omega^*(-\tit'),-\tit',u] \; \vec a_{\mathrm{in}} (\tit').
\end{eqnarray}
The efficiency of this forward quantum memory is thus given by
\begin{eqnarray}\label{eq:rfr}\nonumber
\eta_{{\rm s+fr}} & = & \frac{1}{4\pi^2} \int_0^\infty d\tit_x \int_{-\infty}^\infty du \int_{-\infty}^\infty du'  \int_{-\infty}^0 d\tit  \int_{-\infty}^0 d\tit' \\ 
&& \times \vec a_{\mathrm{in}}^\dag(\tit) \;\mathbb{M}^*[\tilde \Omega^*(-\tit),-\tit,u] \;\mathbb{M}_u^\dag[\tilde \Omega(\tit_x),\tit_x,u] \\ \nonumber
&& \times \mathbb{M}_u[\tilde \Omega(\tit_x),\tit_x,u']  \; \mathbb{M}^T[\tilde \Omega^*(-\tit'),-\tit',u'] \vec a_{\mathrm{in}} (\tit'),
\end{eqnarray}
where $\mathbb{M}_u[\tilde \Omega(\tit),\tit,u] = \mathbb{M}[\tilde \Omega(\tit),\tit,u] e^{iu}$.
Analogously to calculating the efficiencies for storage only, as discussed above, we discretize time and momentum
in order to allow for numerical diagonalization of the large kernel matrix 
$\mathbb{C}_{{\rm s+fr}}[\tit,\tit']  \sim \mathbb{M}^*[\tilde \Omega^*(-\tit),-\tit,u]
 \;\mathbb{M}_u^\dag[\tilde \Omega(\tit_x),\tit_x,u]$
 $\mathbb{M}_u[\tilde \Omega(\tit_x),\tit_x,u']  \; \mathbb{M}^T[\tilde \Omega^*(-\tit'),-\tit',u']/(4\pi^2)$.
 This gives the efficiencies for the forward memory with corresponding incoming light modes $ \vec a_{\mathrm{in}} (\tit)$.

An alternative approach is to use Laplace transform in time
and derive the beam splitter relations in the position and frequency domain.
This is described in detail in Appendix~\ref{app:for}.
The analytical expression for the backward read-out for both solution methods are presented in Appendix~\ref{app:back}.

%%%%%%%%%%%%%%%%%%%%%%%%%%%%%%%%%%%%%%%%%%
\subsection{Results for backward and forward memories}\label{sec:results}
%%%%%%%%%%%%%%%%%%%%%%%%%%%%%%%%%%%%%%%%%%

In the previous subsection as well as in Appendices~\ref{app:for} and~\ref{app:back}, 
we derived the input-output beam splitter relations between the spin-wave and the incoming and outcoming
light modes as well as expressions for the efficiencies for both operation modes (forward and backward).
In this subsection, we present the results of numerical calculations of the maximal efficiencies and the optimized light modes and spin-waves.
In the one-dimensional theory, the optimal storage is the time reversal of the optimal retrieval.
Since the optimal spin-wave excitation is real in that case, the combined process of storage and retrieval 
is independent of detuning~$\tilde \Delta$~\cite{Gorshkov2007a}.
In the three-dimensional theory, however, the spin-wave is not real and one cannot directly use the time reversal argument,
which will result in dependence on the detuning~$\tilde\Delta$. 
Here, for simplicity we only consider the resonant case with $\tilde \Delta=0$.

\begin{figure}[tb]
\begin{center} 
\unitlength 1mm
{\resizebox{87mm}{!}{\includegraphics{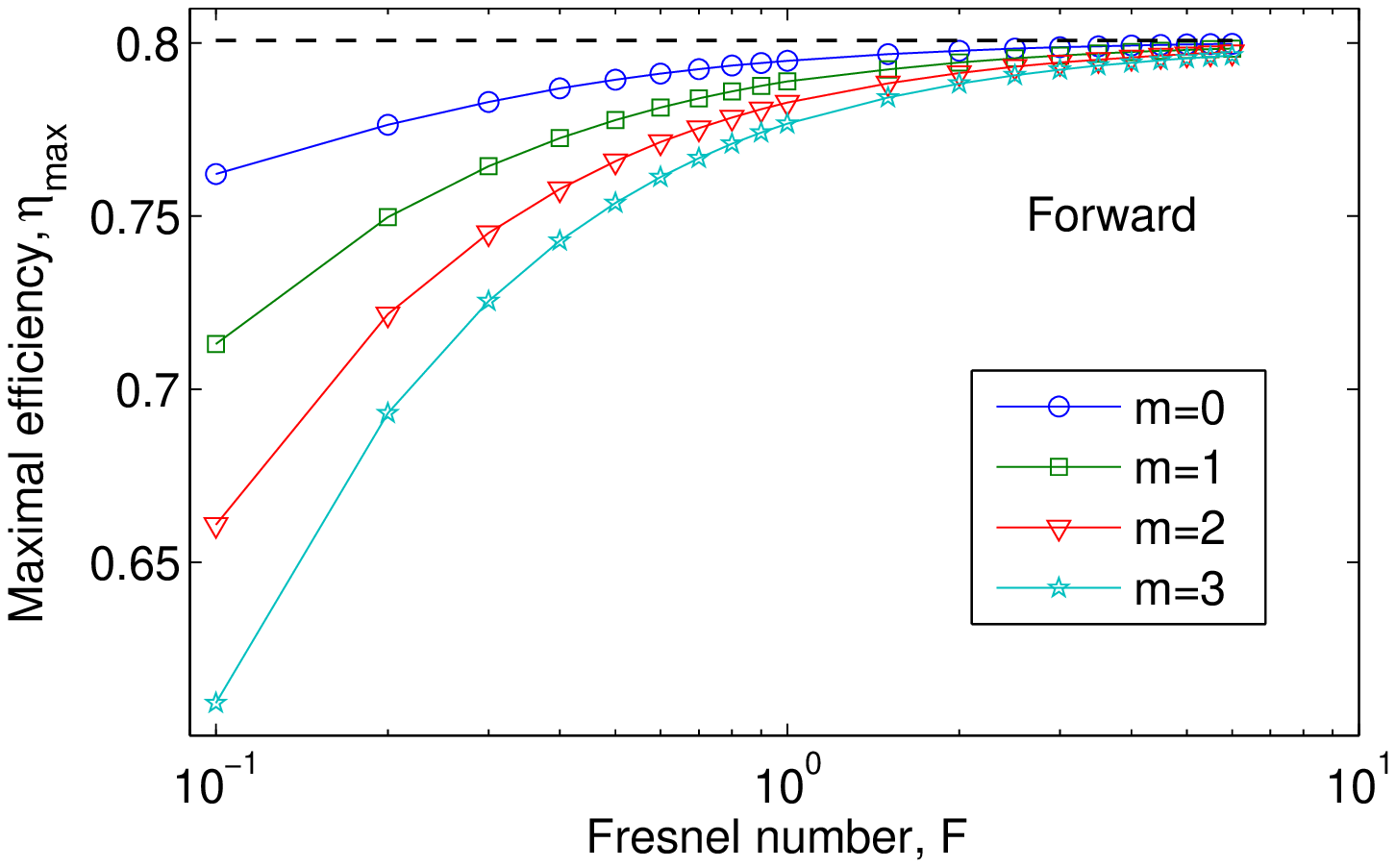}}}
{\resizebox{87mm}{!}{\includegraphics{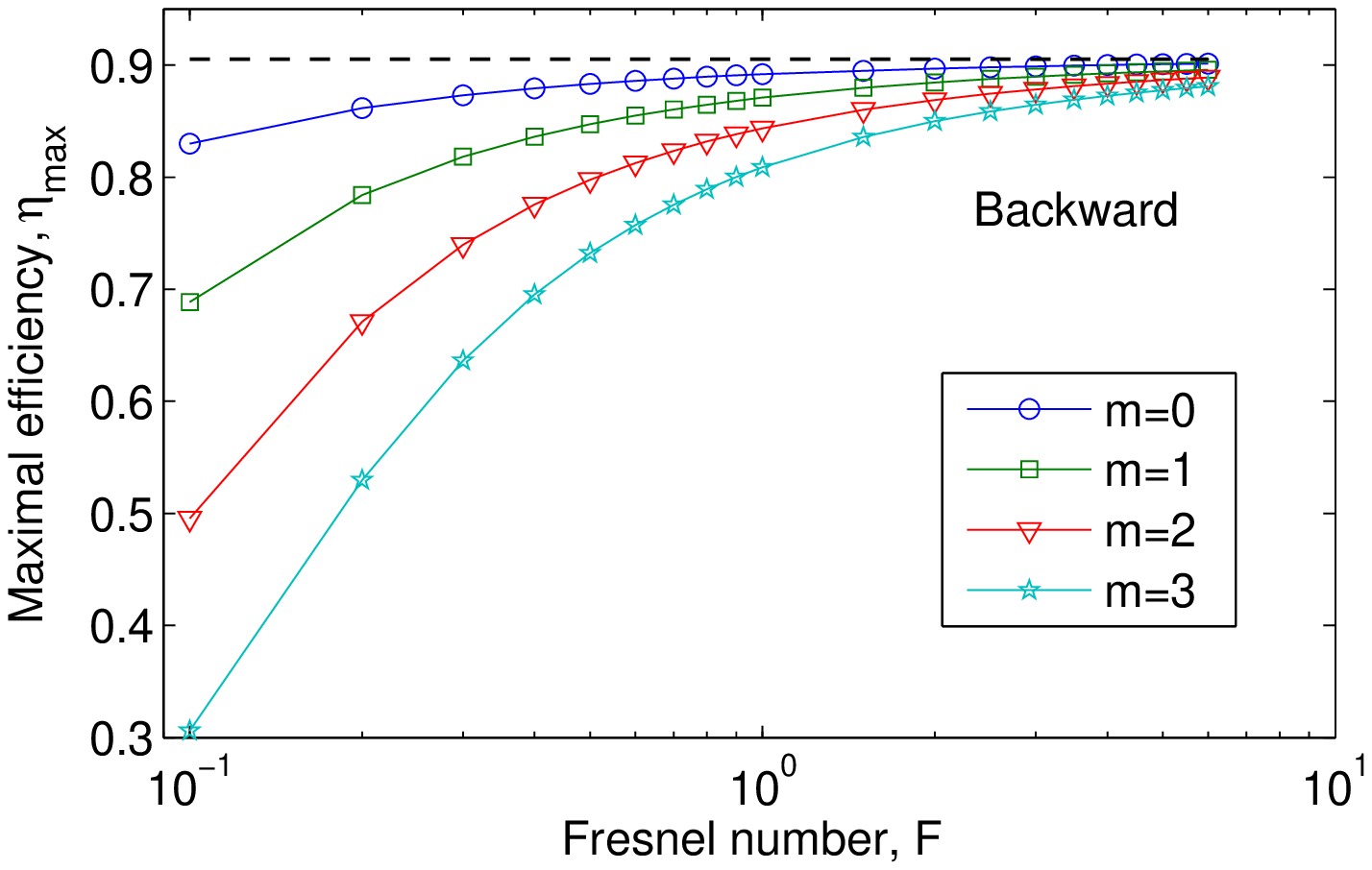}}}
\end{center} 
\caption{\label{fig:forw_d10} (Color online)
The maximal efficiencies for storage followed by read-out in the forward (top) and backward (bottom) direction
as a function of the Fresnel number $F$ for different values of the azimuthal 
quantum number $m$ and for a peak optical depth $d_0=100$. 
The dashed black lines denote the corresponding 1D limits with $F\rightarrow \infty$.}
\end{figure}

In Fig.~\ref{fig:forw_d10}, we show the maximal efficiencies for resonant ($\tilde \Delta=0$) memories operating in the forward (top)
and backward (bottom) direction as a function of the Fresnel number $F$, where the peak optical depth has been fixed to $d_0=100$. 
The highest memory efficiencies for different azimuthal quantum numbers $m=0,1,2$, and~$3$ are shown. 
In addition, for each value of $m$, there are several orthogonal modes with high efficiencies which are not shown in these plots.
For both directions of the read-out, the simulations show the multimode character of the memory, i.e. there are many orthogonal
modes even within each $m$-subspace to store into and retrieve from with high efficiency.
Note that the negative values of the azimuthal quantum number $m$ have the same efficiencies as $|m|$.
The efficiencies decrease for increasing~$m$ due to the spatial light distribution in high $m$ modes.
Light beams with $|m|>0$ vanish towards the center of the sample ($\rho\rightarrow 0$) as $\rho^{|m|}$.
For large $m$, it is therefore harder to focus the input light into the dense center of the atomic ensemble,
which results in lower effective optical depth. Nevertheless, for large Fresnel numbers $F>1$, the efficiencies for successive $m$ decrease only by a few percent in comparison to $m=0$.
For the Fresnel number $F=6$, the efficiency difference between the modes with subsequent $m$ values 
is smaller in the case of forward read-out $\Delta \eta_{\rm for} \approx 0.1 \%$ in comparison to the backward one 
with $\Delta \eta_{\rm back} \approx 0.6 \%$.
We compare the results of the three-dimensional theory with the efficiencies obtained for the one-dimensional memory (dashed black lines). 
The efficiencies approach the 1D limit of $F\rightarrow \infty$ already for small Fresnel numbers $F\approx 1$.
Moreover, the  one-dimensional limit is reached faster for the forward memory.

\begin{figure}[tb] 
\begin{center} 
\unitlength 1mm
{\resizebox{87mm}{!}{\includegraphics{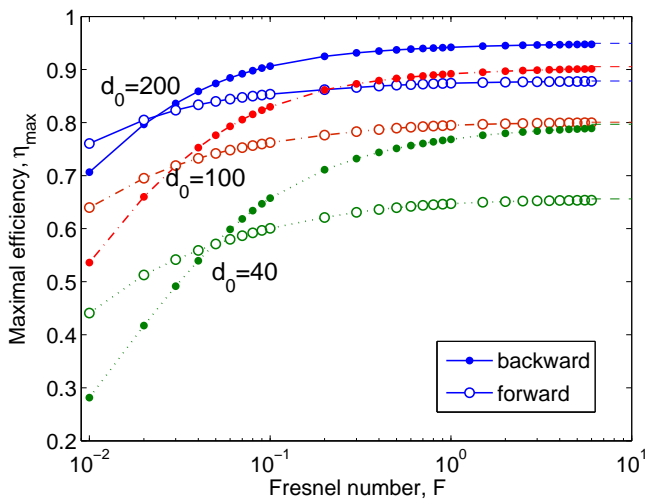}}}
\end{center} 
\caption{\label{fig:comp}(Color online) The maximal efficiencies for storage followed by backward (full circles) and forward (empty circles) read-out
as functions of the Fresnel number $F$ for optical depth $d_0=40$ (bottom, dotted green lines), $d_0=100$ (middle, dashed-dotted red lines),
and $d_0=200$ (top, solid blue lines). The dashed lines for large values of the Fresnel number~$F$ denote the corresponding values
from the one-dimensional theory.}
\end{figure}

In Fig.~\ref{fig:comp}, we compare the highest achievable efficiencies for memories operating in backward (full circles)
and forward (open circles) directions.
The efficiencies are plotted as functions of the Fresnel number $F$ for three values of the peak optical depth 
$d_0 =40, 100$, and~$200$.
The dashed lines for large Fresnel numbers~$F$ denote the corresponding values from the 1D limit.
The one-dimensional theory predicts that the backward memory always yields higher efficiencies~\cite{Gorshkov2007a}.
Here, we find that forward read-out is advantageous for small Fresnel numbers. 
The intersection between the backward and forward efficiencies appears at lower Fresnel numbers
for increasing values of the peak optical depth $d_0$. 
The advantage of forward read-out for small Fresnel numbers $F$ can be understood 
by considering the diffraction of the light beam (in analogy to Gaussian beam diffraction).
In the case of backward retrieval, the $\zt$ dependent part of the phase of the excitation is reversed 
by means of the classical control field with propagation direction opposite to the incoming quantum field.
However, since we are not able to physically complex conjugate the stored spin-wave, the evolution of the transverse phase profile
will be unchanged and will lead to reduced constructive interference of the restored spin-wave:
the output light will accumulate an unwanted spatially varying phase, which in the case of a Gaussian beam 
can be written as $\sim \exp{[i k_\perp r_\perp^2/R(z)]}$. Here, $R(z)$ is the radius of curvature which is proportional to the
Rayleigh range $z_0^2$ as $R(z) \sim z_0^2$. 
In the regime of small Fresnel numbers, the far field divergence angle $\Theta \sim \sqrt{F\lambda/L}$ is small
and the resulting Rayleigh range $z_0 \sim \pi L F$ is short.
Thus for small Fresnel numbers where the diffraction is large, the unwanted phase grows since it scales as $\sim 1/F^2$.
The desired retrieval process is therefore not phase matched and cannot conserve momentum.
This reduces the efficiency of the storage followed by backward retrieval.
In contrast, for the forward retrieval we do not have this problem with the transverse phase profile 
and thus forward read-out is less sensitive to small Fresnel numbers $F$.
This discussed reduction of the backward retrieval efficiency is analogous to the reduction due to accumulated phase for non degenerate 
ground states found in Ref.~[\onlinecite{Surmacz2008}].
In the case of large Fresnel numbers which is well described by the 1D theory, the profile of the excitation is flat in the transverse direction
and no unwanted phase is accumulated. 
Thus for large Fresnel numbers, backward retrieval is favorable. 
For high peak optical depth $d_0$ it is feasible to use only part of the atomic ensemble, whereby
the effective Fresnel number is increased. Therefore the intersection between forward and backward operating memories
for higher peak optical depths $d_0$ occurs at lower values of the Fresnel number $F$.

It is interesting to note that in the case of the DLCZ protocol, where the spin-wave is generated by a parametric gain interaction
and next released from the ensemble with the beam splitter interaction, the above described mode mismatch
(via the unwanted spatially dependent phase accumulation) will occur for the read-out in the forward direction \cite{Andre2005}.
In that case, the backward memory will work better than the forward one for small values of the Fresnel number~$F$.

It was shown in the 1D theory~\cite{Gorshkov2007a}, that in the limit of high optical depths $d_0 \rightarrow \infty$
the inefficiency scales as~$1-\eta_{\rm 1D} \sim 1/d_0$.
We have fitted the minimal inefficiencies in the limit of high values of the Fresnel number $F \leq 1$
with the power dependence on the Fresnel number $1-\eta_{\rm 3D} \sim (1-\eta_{\rm 1D})(1+1/F^l) \sim (1+1/F^l)/d_0$.
For backward operating memories, we obtain $l=3/4$ and for forward retrieval $l=0.9$.

\begin{figure}[tb] 
\begin{center} 
\unitlength 1mm
{\resizebox{87mm}{!}{\includegraphics{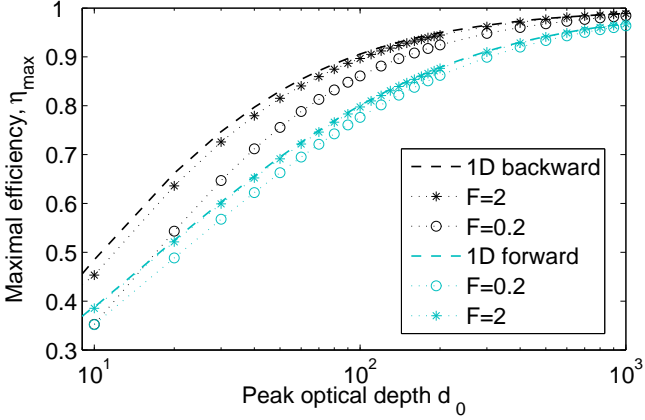}}}
\end{center} 
\caption{\label{fig:d2}(Color online) The maximal efficiencies for the backward (black) and forward (light blue/gray) memory
as a function of the peak optical depth~$d_0$ for $F=0.2$ (empty circles) and $F=2$ (stars). 
The dashed lines correspond to the results of the 1D theory with $F\rightarrow \infty$.}
\end{figure}

Next, we fix the Fresnel number and consider the dependence on the peak optical depth $d_0$.
In Fig.~\ref{fig:d2}, we show the maximal efficiencies for the backward (black) and forward (light blue/gray) memories
for two values of the Fresnel number $F=0.2$ (open circles) and $F=2$ (stars) 
together with the corresponding results in the 1D limit (dashed lines).
For the smaller Fresnel number $F=0.2$ one can see that  for low optical depths~$d_0$ the efficiencies are considerably lower
than the ones corresponding to the 1D limit, which in the case of backward read-out is approached for high~$d_0 \approx 300$. 
The efficiencies for the larger Fresnel number $F=2$ approach the 1D limit already for small optical depths 
especially for the forward operating memory.
The backward and forward efficiencies for $F=0.2$ intersect at the optical depth $d_0 = 10$ 
whereas for larger~$d_0$, the backward memory is advantageous.

The three-dimensional theory presented here allows for calculation of not only the efficiencies but also the optimal light modes to store into 
and retrieve from the memory.
The optimal input and output light modes can be shown to be time reversed complex conjugates of each other 
$\vec a_{\rm{out}}(\tit) = \vec a^*_{\rm{in}}(-\tit)$, which we prove in Appendix~\ref{app:in_out}.
This property can be used for an iterative procedure to obtain the maximum efficiency of the quantum memory.
First, one stores an arbitrary input light mode and retrieve it. In the next steps one uses the time reversed and complex conjugated output light mode
as an input mode for the next run of the experiment. 
Such successive time-reversal iterations  with complex conjugation lead to optimal light storage. 
This is analogous to the proposed procedure in the one-dimensional case~\cite{Gorshkov2007a},
which has been demonstrated experimentally in a warm Rubidium vapor~\cite{Novikova2007}.

The light modes will, in general, be correlated in the spatial and temporal degrees of freedom and the spin-waves in the transverse
and longitudinal spatial modes. This means, for instance, that the transverse spatial profile 
of the optimal incoming light mode would change in time.
Such correlated light modes are disadvantageous from an experimental point of view, where it is simpler to have a 
time independent transverse mode profile.

In order to investigate the degree of correlation for the optimized light modes we have calculated their purities in terms
of the Schmidt decomposition. Any mode can be factorized as
$\vec a_{\rm in}^{\rm \; opt}(\vec r_\perp,\tit) = \sum_n \sqrt{P_n} \; h_n(\vec r_\perp) a_n(\tit)$,
where $h_n(\vec r_\perp)$ and $a_n(\tit)$ depend exclusively on the transverse component and time, 
respectively, and $\sum_n P_n =~1$.
The largest $P_n$ is a measure of the purity of the mode. 
In the ideal pure case, there is only a single component in this expansion which consequently has $P=1$.
In order to determine the purity of the optimized modes, we construct the reduced density matrix
$\varrho(\tit',\tit) = \vec a_{\rm in}^{\rm \; opt\; \dag}(\vec r_\perp,\tit) \vec a_{\rm in}^{\rm \; opt}
(\vec r_\perp,\tit')/\mathrm{Tr}[ \vec a_{\rm in}^{\rm \; opt\; \dag}(\vec r_\perp,\tit) \vec a_{\rm in}^{\rm \; opt}(\vec r_\perp,\tit')]$
the eigenvalues of which are exactly the set $\{P_n\}$.
Analogously, one can calculate the purity for the stored spin-wave excitation which can be decomposed as
$\vec S_{\rm opt} (\vec r_\perp,\zt) = \sum_n \sqrt{P'_n} \; f_n(\vec r_\perp) S_n(\zt)$.

\begin{figure}[tb] 
\begin{center} 
\unitlength 1mm
{\resizebox{87mm}{!}{\includegraphics{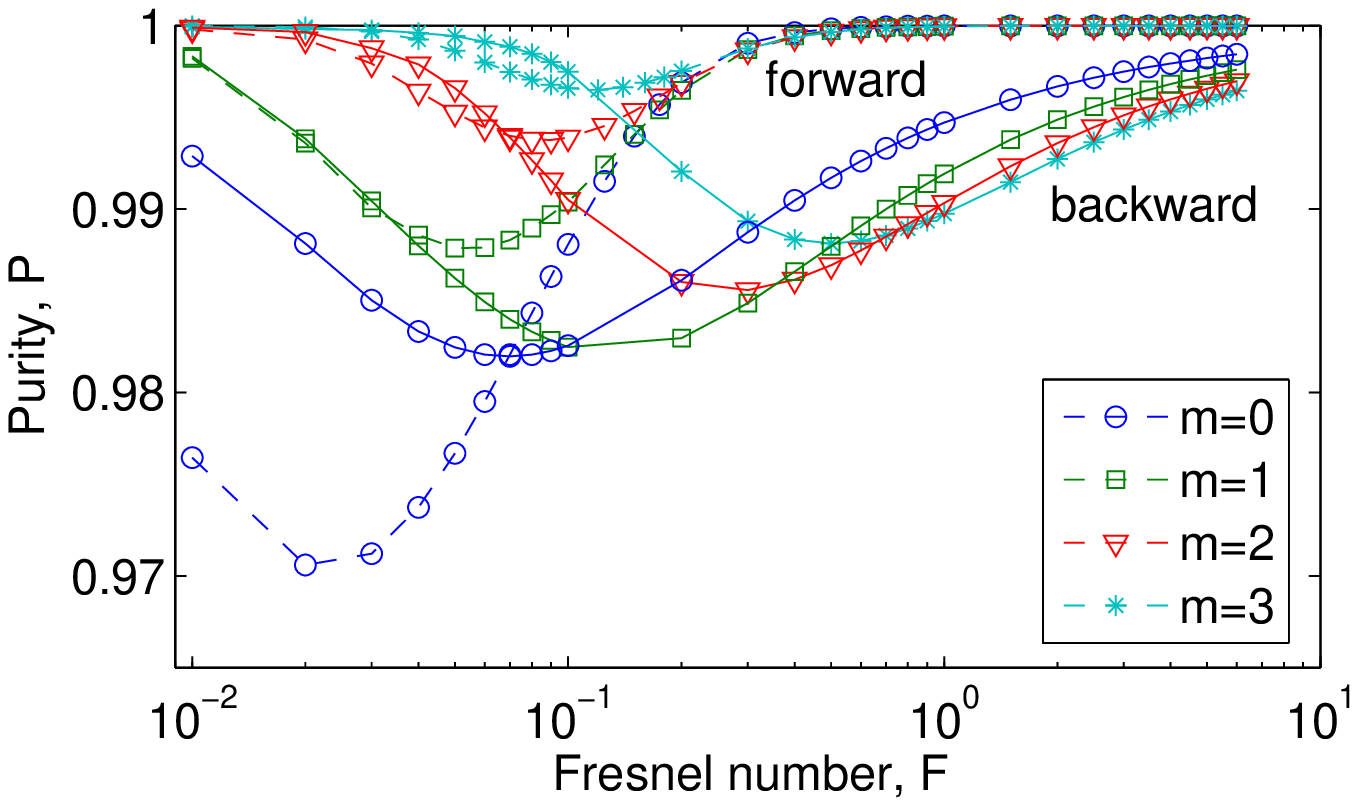}}}
{\resizebox{87mm}{!}{\includegraphics{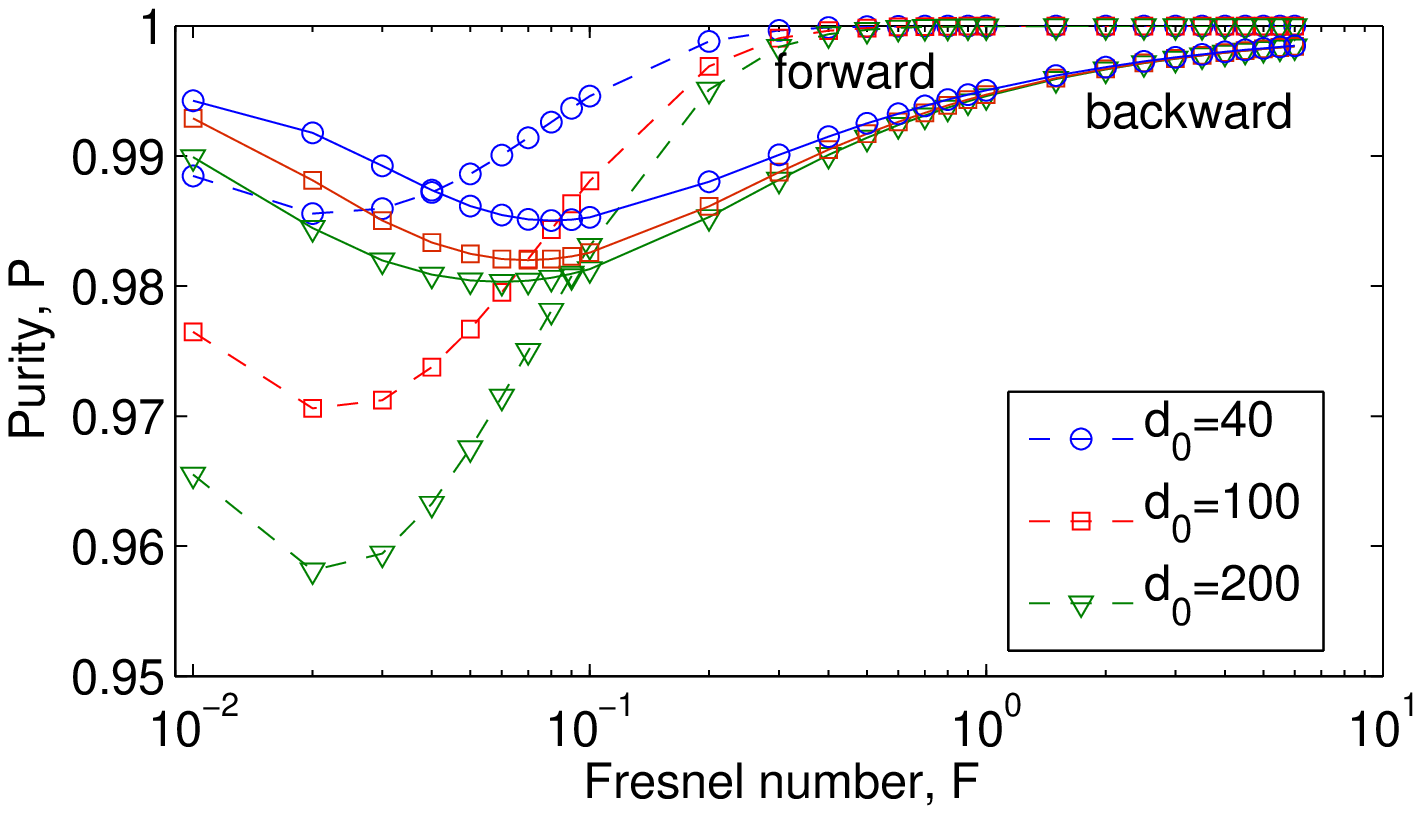}}}
\end{center} 
\caption{\label{fig:pure_m}(Color online)
(top) Purity of the optimal input light mode for backward (solid lines) and forward (dashed lines) memory
as a function of the Fresnel number $F$ for the peak optical depth $d_0 = 100$ and different values of the azimuthal quantum number
$m=0$ (circles), $m=1$ (squares), $m=2$ (triangles), and $m=3$ (stars).
(bottom) The purity of the optimal input light mode for backward (solid lines) and forward (dashed lines) memory
as a function of the Fresnel number $F$ for $m=0$ and different values of the peak optical depth $d_0 = 40$ (circles),
$d_0=100$ (squares), and $d_0=200$ (triangles).}
\end{figure}

In Fig.~\ref{fig:pure_m} (top), the purity of the optimal input light mode for various azimuthal quantum numbers $m$
is shown as a function of the Fresnel number~$F$ for
a peak optical depth $d_0=100$ for both forward (dashed lines) and backward (open lines) operating quantum memories. 
The purities of the optimal output and input light modes are equal since the output light is the time
reversed complex conjugate of the input mode (see Appendix~\ref{app:in_out}).
First of all, the purities are in general very close to one.
For large values of the Fresnel number~$F$, the purity is higher in the case of the forward memory
even though the spin-wave is not factorizable.
It is interesting that in all cases, there exists a minimum of the purity which depends on $m$.
If one decreases the Fresnel number, the purity drops at most by a few percent and is bounded from below at $\sim 96\%$.
Next, we investigate how the optical depth influences the purity, see Fig.~\ref{fig:pure_m} (bottom).
For large~$F$, the purities have the same value for all three values of the peak optical depths $d_0 = 40, 100$, and $200$.
The situation changes for small $F$, where the purity decreases more for large values of $d_0$. 
Again, in this regime, the purity for the forward read-out decreases to lower values than the backward one. 
All in all, the purities of the optimal light pulses are close to one and their dependence on the physical parameters $d_0$ and $F$
as well as the choice of the azimuthal quantum number $m$ is very weak, which is advantageous for experimental realizations. 
One can, however, estimate how the fact that one uses only the dominant Schmidt component of the optimal input light pulse 
and measures only the Schmidt component of the corresponding output light will affect the efficiencies.
The efficiency of the full optimized light $\eta$ will at most be reduced to $\eta_{\rm pure} \approx \eta(1-4\varepsilon)$, 
where $\varepsilon=1-P$ is the lack of purity (see Appendix~\ref{app:purity} for derivation). 

\begin{figure}[tb] 
\begin{center} 
\unitlength 1mm
{\resizebox{87mm}{!}{\includegraphics{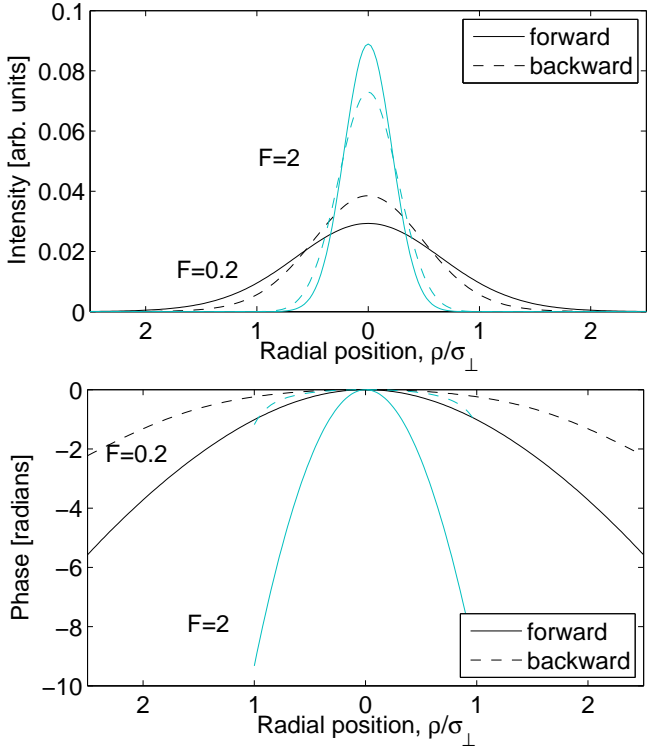}}}
\end{center} 
\caption{\label{fig:in}(Color online)
(top) The spatial profile of the intensity of the optimal incoming light pulse for $F=0.2$ (black lines)
and $F=2$ (light blue/gray lines) for backward (dashed lines) and forward (solid lines) memory ($d_0=100$).
(bottom) The corresponding spatial profile of the phases as in (top).}
\end{figure}

Given the fact that the optimal light modes have very high purities, it is reasonable to extract the dominant Schmidt component 
$h_n(\vec r_\perp)$.  
In Fig.~\ref{fig:in} (top), we plot the spatial intensity profile of the optimized incoming light pulses for small 
$F=0.2$ (black lines) and large $F=2$ (light blue/gray lines)
Fresnel numbers, for backward (dashed lines) and forward (solid lines) memory.
The transverse spatial shape of the optimal light pulses with the azimuthal quantum number $m=0$ is well approximated by a Gaussian beam.
For atomic ensembles with large Fresnel number, the optimized light is more localized in comparison to the sample size than for small~$F$.
In the latter case, too narrow pulses would entail a strong diffraction leading to excitation leaking out of the sample 
and weaker interaction with atoms. 
Thus, larger values of the Fresnel number allow for more localized light pulses, which in consequence will lead to a larger number
of available modes with high efficiencies.
The full multimode capacity of spatial memories based on the $\Lambda$-type atomic ensembles will be studied in detail 
elsewhere~[\onlinecite{Grodecka-Grad2011}].
The spatial profiles of the intensity for backward and forward memory are only slightly different. The width of the pulse
is smaller for the backward read-out for small~$F$ and larger for large~$F$ due to diffraction. 

We also show the spatial phase profiles of the optimal input light, see Fig.~\ref{fig:in} (bottom). 
The lines correspond to the discussed intensities shown in Fig.~\ref{fig:in} (top). 
One can see, that for both small and large $F$, the phase profile for backward memory is flatter than the one for the forward read-out.
It results from the fact, that the focal plane with a flat phase profile of the optimal stationary excitation for the forward read-out 
lies in the middle of the sample at $\zt=0.5$
while for the backward case it will be placed much closer to the read-out end of the ensemble.
Thus the propagation length is longer for the forward retrieval and the accumulated phase is larger. 

\begin{figure}[tb]
\begin{center} 
\unitlength 1mm
{\resizebox{87mm}{!}{\includegraphics{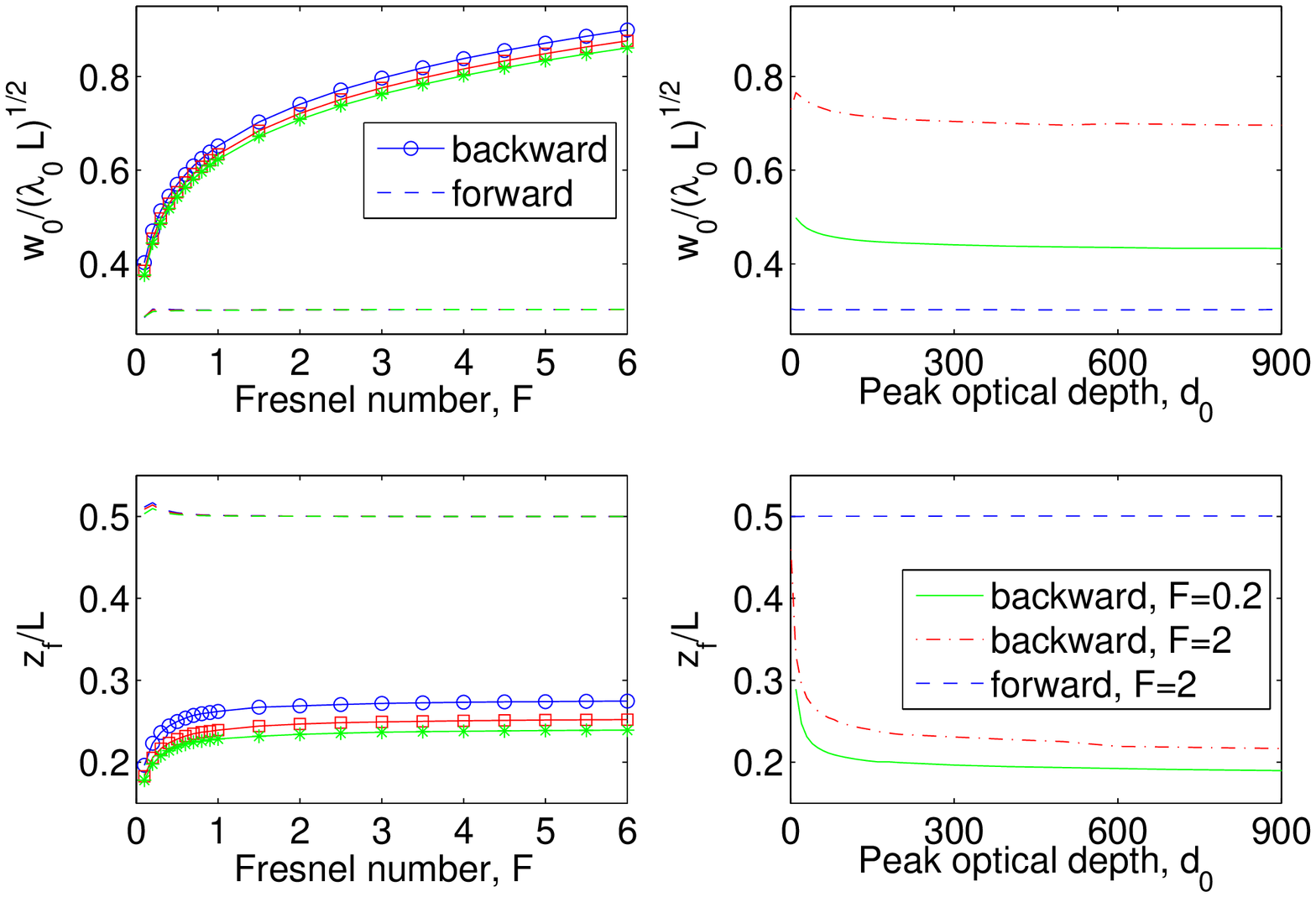}}}
\end{center} 
\caption{\label{fig:waist}(Color online)
(top, left) The dimensionless waist of the optimal input light multiplied by the square root of the Fresnel number~$F$,
$\tilde w_0 \sqrt{F} = w_0/(\lambda_0 L)$,  
as a function of the Fresnel number for backward (solid lines with markers) and forward memory (dashed lines).
Here, we present the fits for the peak optical values of $d_0 = 40$ (blue circles), $d_0 = 100$ (red squares), and $d_0 = 200$ (green stars).
(top, right) The dimensionless waist of the optimal input light scaled as in (top, left)
as a function of the peak optical depth~$d_0$ for forward read-out with $F=2$ (dashed blue line) and for backward 
read-out with $F=0.2$ (solid green line) and $F=2$ (dashed-dotted red line). 
(bottom, left) The dimensionless distance between the beginning of the atomic
ensemble and the focal plane in the free space $\tilde z_{\rm f} = z_{\rm f}/L$ of the optimal input light mode as a function of the Fresnel number~$F$.
(bottom, right) The distance $\tilde z_{\rm f}$ as a function of the peak optical depth~$d_0$.}
\end{figure}

The information gained from the spatial profiles of the intensity and phase allows us to calculate
the beam waist of the optimal input light mode and the distance from the beginning of the ensemble
to the focal plane for the light propagating without the atoms. 
We fit the dominant Schmidt component of the optimal mode with a Gaussian beam
$h_{\rm pure}(\tilde\rho) \sim \exp[-\tilde\rho^2/\tilde w^2(\zt))]\exp[i\tilde k \tilde\rho^2/(2\tilde R(\zt))]$.
Here, $\tilde w(\zt) = \tilde w_0 \sqrt{1+\zt^2/\zt_0^2}$ is the dimensionless spot size
and $\tilde w(\zt) = w(\zt)/\sigma_\perp$, $\tilde w_0 = w_0/\sigma_\perp$ is the dimensionless beam waist,
$\tilde \rho = \rho/\sigma_\perp$, and $\tilde k = k L$.
The phase profile of the input mode is fit to find the dimensionless radius of curvature $\tilde R(\zt) = \zt + \zt_0^2/\zt$.
From these we calculate the beam waist 
\begin{equation}
\tilde{w}_{0}=\sqrt{\pi^{2}\frac{\tilde{w}^{2}(\zt_{\mathrm f})}{\tilde{R}^{2}(\zt_{\mathrm f})}+\frac{1}{\tilde{w}^{2}(\zt_{\mathrm f})}},
\end{equation}
and the distance of the focal plane from the beginning of the sample
\begin{equation}
\tilde{z}_{\text{f}} = -\frac{ \pi^2  \tilde w^4 F \tilde{R}(\zt_{\mathrm f})}{ \pi^2  \tilde w^4 + \tilde{R}^{2}(\zt_{\mathrm f})}.
\end{equation}
In order to check the quality of the fitting procedure, we calculated the overlap between the 
pure part of the optimal light mode and the fitted mode. These overlaps are equal or larger
that $0.996$ and $0.994$ for forward and backward operating memory, respectively. 

In Fig.~\ref{fig:waist} (top, left), we plot the dimensionless waist of the pure part of the input light beam
multiplied by the square root of the Fresnel number $F$, $\tilde \omega_0 \sqrt{F} = \omega_0/\sqrt{\lambda_0 L}$, 
as a function of the Fresnel number. 
For the forward operating memory (dashed lines), we can see a universal behavior that the waist
is independent of the Fresnel number, it is a constant equal to $w_0 = 0.3\sqrt{\lambda_0 L} \approx 1/(\pi\sqrt{\lambda_0 L})$,
given by the wavelength $\lambda_0$ and the length of the sample $L$.
For the optimal light beam, it is desirable to both have equal waist along the entire atomic ensemble to provide high collective interaction
with atoms and to be as focused as possible.
The trade-off between these effects is achieved for an optimal waist which has a Rayleigh range $z_0 \approx L$ independent of~$F$.
The situation is different for the backward read-out (solid lines with markers), where the spin-wave excitation is not symmetric in the~$\zt$ direction.
Here, the optimal beam waist grows for increasing Fresnel number. 
The dependence of the beam waist on the peak optical depth is shown in Fig.~\ref{fig:waist} (top, right).
One can see that it is again constant for the forward read-out whereas for the backward one
it decreases slowly with growing optical depth $d_0$.

The corresponding distance $\zt_{\rm f}$ between the focal plane and the beginning of the ensemble is shown in Fig.~\ref{fig:waist} (bottom, left)
first as a function of the Fresnel number~$F$. 
For the forward read-out, the focal plane is always in the middle of the sample ($\zt = 0.5$).
In the backward operating memory, the focal plane is closer to the beginning of the ensemble and decreases only slightly
for small Fresnel numbers $F\leq 1$. For large values of $F$ it reaches a constant value depending on the peak optical depth.
The influence of the optical depth on the distance $\zt_{\rm f}$ is shown in Fig.~\ref{fig:waist} (bottom, right).
Again, for the forward operating memory, it is constant and equals to $\zt_{\rm f} = 0.5$.
In the case of backward retrieval, high optical depths allow for focusing closer to the beginning of the ensemble.
The distance increases for decreasing peak optical depth $d_0$ as the spin-wave becomes more and more symmetric
due to the increasing transparency of the medium. 

\begin{figure}[tb]
\begin{center} 
\unitlength 1mm
{\resizebox{87mm}{!}{\includegraphics{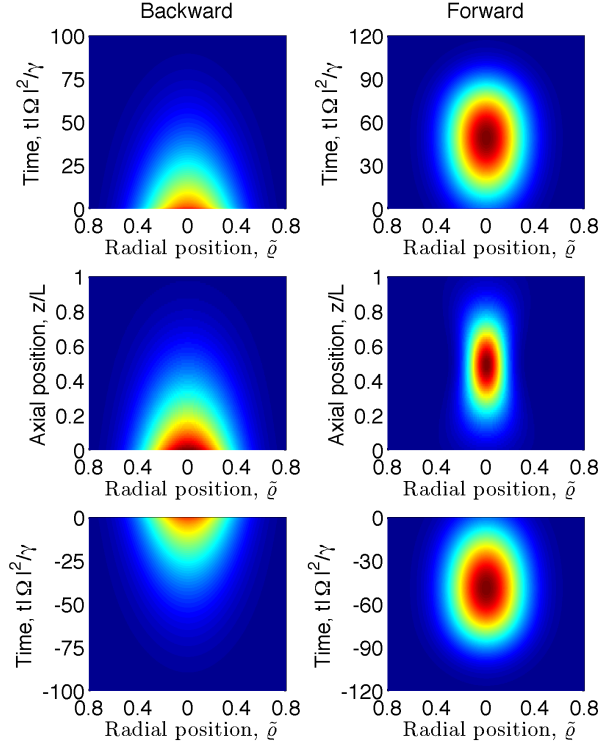}}}
\end{center} 
\caption{\label{fig:set}(Color online) (top) The spatial and temporal shape of the optimal outcoming light pulse.
(middle) The optimal spin-wave excitation in the atomic ensemble. (bottom) The corresponding incoming light mode.
(left) The backward memory with read-in and read-out at $\tilde z = 0$. 
(right) The forward memory. All results are for $F=2$ and $d_0=100$.
The radial position is scaled with the atomic ensemble width as $\tilde\rho = \rho/\sigma_\perp$ and time according to
$\tit = t |\Omega|^2/ \gamma$.}
\end{figure}

Knowing the spatial profile of the intensity and phase of the optimized incoming light mode,
it is also interesting to study the spatial profiles of the corresponding spin-waves 
as well as the spatio-temporal shapes of the outgoing light modes.
In Fig.~\ref{fig:set}, we show the full set of light modes and spin-wave excitations for the backward (left column)
and forward operating memory (right column) with Fresnel number $F=2$ and with peak optical depth $d_0=100$.
We assumed for simplicity a constant Rabi frequency~$\tilde \Omega$ and the memory is resonant~$\tilde \Delta = 0$.
In the case of the backward memory, it is preferable (especially for high $d_0$ \cite{Gorshkov2007a}) to store most of the excitation 
closer to the beginning of the atomic ensemble ($\zt = 0$) where the read-in and read-out are carried out, see Fig.~\ref{fig:set} (middle, left). 
Thus, the light does not need to propagate through the entire ensemble, which leads to reduced decoherence due to spontaneous emission
as well as there being less irreversible phase acquired. 
Such a spin-wave is obtained with an incoming light mode, where the time dependence is a "cut/half" Gaussian with the intensity maximum
at the end of the write-in process, see Fig.~\ref{fig:set} (bottom, left).
This is in agreement with the results of the one-dimensional theory~\cite{Gorshkov2007a}
and has been also demonstrated experimentally~\cite{Novikova2007}.
For the forward memory, Fig.~\ref{fig:set} (middle, right), it is advantageous to store a symmetric and smooth spin-wave,
which is achieved with a light pulse with a Gaussian type time dependence (with a very small final cut-off),
see Fig.~\ref{fig:set} (bottom, right).
As mentioned earlier, the outgoing light modes are the time reversals with respect to the input light modes
$\vec a_{\rm{out}}(\tit) = \vec a^*_{\rm{in}}(-\tit)$ with the complex conjugated spatial profiles.
If the memory is operated off-resonantly, the optimal light modes and spin-waves may have different transverse profiles
from the ones presented here, since both the absorption and refraction during light propagation will be affected.

%%%%%%%%%%%%%%%%%%%%%%%%%%%%%%%%%%%%%%%%%%
\section{Conclusion}\label{sec:concl}
%%%%%%%%%%%%%%%%%%%%%%%%%%%%%%%%%%%%%%%%%%

We have solved the three-dimensional problem of light storage in a $\Lambda$-type atomic ensemble of finite spatial extent.
Specifically, we considered cylindrically symmetric samples with a constant atom density along the axial direction~$\zt$.
We have shown that within the paraxial approximation, the only two important physical parameters 
characterizing the memory are the peak optical depth $d_0$ and the Fresnel number $F$ of the atomic ensemble. 
The maximal efficiencies for retrieval only as well as the full process of storage followed by retrieval have been calculated.
We considered both the forward and the backward operation modes of the memories
and showed that both lead to high efficiencies for dense atomic media $d_0 \geq 40$
already for small Fresnel numbers $F \gtrsim 0.1$.
For small Fresnel numbers it is favorable to operate in the forward direction,
whereas for large~$F$, backward read-out yields higher efficiencies.
The optimal incoming modes and corresponding stored spin-waves and output light modes have been presented for a set of physical parameters
$F=2$ and $d_0 = 100$ for both backward and forward operating memory. 
The purity values of the optimal incoming light modes are close to one, which is advantageous for experimental realizations.
Moreover, we calculate the beam waist and the distance between the focal plane and the beginning of the ensemble
for the optimal input light modes. 

The results presented here are of direct importance for the experiments currently performed, 
where both the optical depth $d_0$ and the Fresnel number~$F$
are limited by practical constraints. It may serve as a guide for choosing the optimal input light pulses to store given a
set of physical parameters $F$ and $d_0$, which may be constrained by, for instance, 
the limit in the achievable density (optical depth) of cold atoms or imperfection of the initialization by the optical pumping. 
Thus the knowledge of which light modes have optimal efficiency for such quantum memories is crucial~\cite{Reim2010}.

%%%%%%%%%%%%%%%%%%%%%%%%%%%%%%%%%%%%%%%%%%
\section{Acknowlegments}
%%%%%%%%%%%%%%%%%%%%%%%%%%%%%%%%%%%%%%%%%%

This work was supported by the European Project HIDEAS (FP7-ICT-221906)
and the Marie Curie Research Training Network EMALI (MRTN-CT-2006-035369).
We thank A.~Griesmaier, N.~S.~Kampel, J.~H.~M{\"u}ller, and E.~S.~Polzik for fruitful discussions
and F.~Kaminski for reading the manuscript. 

\begin{appendix}
%%%%%%%%%%%%%%%%%%%%%%%%%%%%%%%%%%%%%%%%%%
\section{Proof for optimal read-out}\label{app:proof}
%%%%%%%%%%%%%%%%%%%%%%%%%%%%%%%%%%%%%%%%%%

In this  Appendix, we present the proof that the read-out efficiency is independent of the detuning $\tilde \Delta$
and the temporal shape of the control field $\tilde \Omega(\tit)$.

We apply the spatial Laplace transform to the equations of motion~(\ref{eq:a})-(\ref{eq:s}) 
and we find that the relation in Eq.~(\ref{eq:ppp}) corresponds to the fact, that each matrix $\mathbb{A}(u',u)$ 
should obey the Sylvester matrix equation
\begin{eqnarray}\label{eq:syl}
\left(\frac{1}{2} + \mathbb{B} \frac{\frac{1}{4}d_0}{u'^* - i\frac{\vec k_\perp^2 \sigma_\perp^2}{4\pi F} }  
\mathbb{B} \right)  \mathbb{A}(u',u) \phantom{asdfasdf} \\ \nonumber
+  \mathbb{A}(u',u) \left(\frac{1}{2} +  \mathbb{B} \frac{\frac{1}{4}d_0}{u +i\frac{\vec k_\perp^2 \sigma_\perp^2}{4\pi F} } 
 \mathbb{B} \right) = - \mathbb{I},
\end{eqnarray}
where $\mathbb{I}$ is the unity matrix and the coupling matrix $\mathbb{B}$, Eq.~[\ref{eq:bb}], has been written 
in the compact matrix notation.
We can write this equation as $ \mathbb{L}  \mathbb{A} + \mathbb{A} \mathbb{R} = -\mathbb{I}$
with $ \mathbb{L} = \frac{1}{2} + \frac{1}{4}d_0 \mathbb{B}(u'^* - i\frac{\vec k_\perp^2 \sigma_\perp^2}{4\pi F})^{{-1}} \mathbb{B}$
and $ \mathbb{R} = \frac{1}{2} + \frac{1}{4}d_0  \mathbb{B} (u +  i\frac{\vec k_\perp^2 \sigma_\perp^2}{4\pi F})^{{-1}}  \mathbb{B}$.
It has a solution when the eigenvalues of the two matrices $ \mathbb{L}$ and $ \mathbb{R}$ obey: 
$\mathrm{EIG}(\mathbb{L}) \neq -\mathrm{EIG}(\mathbb{R})$~\cite{bartels1972},
where ${\rm EIG}$ denotes the set of eigenvalues. 
We are only interested in purely imaginary $u$ and $u'$, where $u=iy$ and $u'=iy'$, then Eq.~(\ref{eq:syl}) reads
\begin{eqnarray}
\left(\frac{1}{2} +   \mathbb{B} \frac{\frac{1}{4}i d_0 }{y' - \frac{\vec k_\perp^2 \sigma_\perp^2}{4\pi F} }  \mathbb{B} \right)    
 \mathbb{A}(u',u) \phantom{asdfadf} \\ \nonumber
+  \mathbb{A}(u',u) \left(\frac{1}{2} -  \mathbb{B} \frac{\frac{1}{4}id_0}{y + \frac{\vec k_\perp^2 \sigma_\perp^2}{4\pi F} } 
 \mathbb{B}\right)  = -  \mathbb{I}.
\end{eqnarray}
The coupling matrix $ \mathbb{B}$ is real and symmetric. Thus, one can see that the eigenvalues of the matrix $\mathbb{L}$ fulfill
\begin{eqnarray}\nonumber
{\rm EIG}  ( \mathbb{L})= {\rm EIG} \left(\frac{1}{2} +  \mathbb{B} \frac{\frac{1}{4}i d_0 }{y' - \frac{\vec k_\perp^2 \sigma_\perp^2}{4\pi F} }  
 \mathbb{B} \right)    =  \frac{1}{2} + i {\rm EIG} ( \mathbb{W'}),
 \end{eqnarray}
where $  \mathbb{W'} =  \frac{1}{4} d_0 \mathbb{B} \left(y' - \frac{\vec k_\perp^2 \sigma_\perp^2}{4\pi F} \right)^{-1}  \mathbb{B}$ 
is a real and symmetric matrix. 
On the other hand the eigenvalues of the matrix $ \mathbb{R}$ can be written as
${\rm EIG} ( \mathbb{R}) = \frac{1}{2} - i {\rm EIG }( \mathbb{W})$ with a real matrix
$ \mathbb{W} =  \frac{1}{4} d_0  \mathbb{B} \left(y + \frac{\vec k_\perp^2 \sigma_\perp^2}{4\pi F} \right)^{-1}  \mathbb{B}$.
Thus, the real parts are always equal, $\mathrm{Re}[{\rm EIG} ( \mathbb{L})] = \mathrm{Re}[{\rm EIG} ( \mathbb{R})] =\frac{1}{2}$.
In consequence, the condition $\mathrm{EIG}( \mathbb{L}) \neq -\mathrm{EIG}( \mathbb{R})$ always holds,
which proves that there is a solution to the Sylvester equation~(\ref{eq:syl}).
This means that one can rewrite the position dependent loss and thus the read-out efficiency in terms of only the 
matrix $\mathbb{A}(u',u)$ and the spin wave $\vec S(u,0)$, which only depend on the optical depth $d_0$
and the Fresnel number $F$ while being independent of the detuning~$\tilde \Delta$
and the temporal shape of the control field~$\tilde \Omega(\tit)$.

%%%%%%%%%%%%%%%%%%%%%%%%%%%%%%%%%%%%%%%%%%
\section{Alternative approach to forward read-out}\label{app:for}
%%%%%%%%%%%%%%%%%%%%%%%%%%%%%%%%%%%%%%%%%%

In Section~\ref{sec:forw}, we presented the analytical expressions for the efficiencies of the storage followed by forward read-out,
where Laplace transform in space was employed. 
In this Appendix, we present an alternative approach, where we perform Laplace transform in time,
which for the storage procedure is defined by ${\cal L}\{g(\tit)\}=\int_{0}^{\infty}e^{-\tilde \omega \tit}g(\tit)d\tit$,
where $\omt = \omega/\gamma$ is the dimensionless frequency. 
For the retrieval, the integral limits are from $-\infty$ to $0$.
Here, for simplicity we assume the driving field to be constant in both space and time 
$\tilde \Omega(\vec r,\tit) = \tilde \Omega$ (see discussion in Secs.~\ref{sec:model} and \ref{sec:forw})
and perform the derivation also in the co-moving frame.
The input-output relations between the light modes and the stationary excitations then read 
\begin{eqnarray}
\vec{a}_{\text{out}}(\omt) & = &\int_{0}^{1}d\tilde{z}\; \mathbb{K}[\tilde \Omega,\omt,\tilde{z}]\; \vec{S}_0(\tilde{z}),\\
\vec{S}_0(\tilde{z}) &= &\frac{1}{2\pi i}\int_{-i\cdot\infty}^{i\cdot\infty}d\omt \mathbb{K}^{\text{T}}
[\tilde \Omega^{*},\omt,1-\tilde{z}]\vec{a}_{\text{in}}(\omt).
\end{eqnarray}
Here, the matrices are
\begin{eqnarray*}
\mathbb{K}[\tilde \Omega,\omt,\tilde{z}] & = & e^{\mathbb{E}(\omt)\cdot(1-\tilde{z})}\mathbb{H}(\omt),\\
\mathbb{H}(\omt) & = & -\frac{\sqrt{d_0} \; \tilde \Omega }{4\omt (i\tilde \Delta + \frac{1}{2}) + |\tilde \Omega|^2} \mathbb{B},\\
\mathbb{E}(\omt) & = & - i \frac{\vec k_{\perp}^{2} \sigma_\perp^2}{4\pi F} 
- \frac{d_0 \omt}{4\omt (i\tilde \Delta + \frac{1}{2}) + |\tilde \Omega|^2} \mathbb{B}^2.
\end{eqnarray*}

If we normalize the incoming light mode according to $\frac{1}{2\pi i}\int_{-i\cdot\infty}^{i\cdot\infty}d\omt |\vec{a}_{\text{in}}(\omt)|^{2}=1$,
the storage efficiency is given by
\begin{eqnarray}
\eta_{\text{s}} & = & \frac{1}{4\pi^2} \int_0^1 d \zt \int_{-i\cdot\infty}^{i\cdot\infty}d\omt \int_{-i\cdot\infty}^{i\cdot\infty}d\omt' 
\vec a_{\text{in}}^\dag(\omt) \\ \nonumber
&& \times \mathbb{K}^*[\tilde \Omega^*, \omt, 1-\zt]  \mathbb{K}^T[\tilde \Omega^*, \omt', 1-\zt]  \vec a_{\text{in}}(\omt').
\end{eqnarray}

The relation between the incoming and outcoming light modes is
\begin{eqnarray}
\vec a_{\text{out}}(\omt) & = & \frac{1}{2\pi i} \int_0^1 d \zt \int_{-i\cdot\infty}^{i\cdot\infty}d\omt'
\mathbb{K}[\tilde \Omega, \omt, \zt] \\ \nonumber && \times \mathbb{K}^T[\tilde \Omega^*, \omt', 1-\zt]  \vec a_{\text{in}}(\omt')
\end{eqnarray}
and thus, the combined efficiency for storage followed by forward retrieval is
\begin{eqnarray} \nonumber
\eta_{\text{s+fr}} & = & \frac{1}{8\pi^3 i} \int_{-i\cdot\infty}^{i\cdot\infty}d\omt \int_0^1 d \zt \int_0^1 d \zt'  
\int_{-i\cdot\infty}^{i\cdot\infty}d\nu \int_{-i\cdot\infty}^{i\cdot\infty}d\nu' \\
&& \times \vec a_{\text{in}}^\dag(\nu)  \mathbb{K}^*[\tilde \Omega^*, \nu, 1-\zt]  \mathbb{K}^\dag[\tilde \Omega, \omt, \zt] \\ \nonumber
&& \times \mathbb{K}[\tilde \Omega, \omt, \zt']  \mathbb{K}^T[\tilde \Omega^*, \nu', 1-\zt'] \vec a_{\text{in}}(\nu').
\end{eqnarray}
Analogously to the procedure discussed below Eqs.~(\ref{eq:etas}) and (\ref{eq:rfr}), we diagonalize the kernel matrix 
$\sim \mathbb{K}^*[\tilde \Omega^*, \nu, 1-\zt]  \mathbb{K}^\dag[\tilde \Omega, \omt, \zt] 
\mathbb{K}[\tilde \Omega, \omt, \zt']  \mathbb{K}^T[\tilde \Omega^*, \nu', 1-\zt']$ in order to find
the eigenvalues giving the efficiencies of the storage followed by forward read-out. The corresponding eigenvectors
represent the optimized set of incoming light modes~$\vec a_{\text{in}}(0,\nu)$.

%%%%%%%%%%%%%%%%%%%%%%%%%%%%%%%%%%%%%%%%%%
\section{Read-out in the backward direction}\label{app:back}
%%%%%%%%%%%%%%%%%%%%%%%%%%%%%%%%%%%%%%%%%%

In this Appendix, storage followed by backward retrieval is analyzed. 
In the 1D model, the optimal spin-wave for retrieval is real and the optimal storage efficiency follows from the optimal read-out efficiency
by means of a time reversal argument as simply $\eta_{\rm r} = \eta_{\rm s}$. 
For the three-dimensional problem, the optimal spin-wave is in general complex and in order to use the same argument,
one would have to physically perform complex conjugation of the spin-wave, which is impossible. 
Thus, we need to optimize the full procedure as in the case of the forward memory.

After we store the light mode in the spin-wave $\vec S_0(\zt)$ we want to retrieve it with a control field 
propagating in the opposite direction to the one used for storage. One way to solve this problem is to rewrite 
the equations of motion~(\ref{eq:aa}) and~(\ref{eq:pa}) for the opposite propagation direction ($-\zt$).
On the other hand, one can also flip the spin-wave according to $\vec S_0(\zt) \rightarrow \vec S_0(1-\zt)$
and next perform the read-out in the forward direction.
In the Laplace transformed spatial variables, 
this transformation translates $\vec S_0(u)$ into $\vec S_0(-u)e^{-iu}$. 

Note that this method of describing backward read-out by the equations for forward propagation
is only valid for degenerate ground states $|0\rangle$ and $|1\rangle$.
Otherwise, the momentum conservation would be broken and the proper spin-wave to retrieve from in the backward direction
would read $\vec S_0(1-\zt) e^{2i\Delta \tilde k \zt}$ with $\Delta \tilde k = L (\omega_0-\omega_1)/c$.
This accumulated phase leads to a decrease in the efficiency of the memory operating in the backward direction~\cite{Gorshkov2007a}.
In Ref.~[\onlinecite{Surmacz2008}], it was proposed to partially overcome this problem by phase matching with signal and control pulses that are not collinear.

In contrast to the forward memory, one needs to transform the spin-wave back to position space $\zt$ after the storage is completed,
otherwise the read-out is carried out not only from the atomic ensemble but also from the leaked excitation with $\zt>1$.
The input-output beam splitter relations for the backward read-out are
\begin{eqnarray}\nonumber
\vec a_{\mathrm{out}}(\tit) & = & \frac{1}{2\pi i} \int_{-\infty}^\infty du \int_0^1 d\zt e^{-iu\zt} 
\; \mathbb{M}[\tilde \Omega(\tit),\tit,-u]  \; \vec S_0(\zt),
\end{eqnarray}
\begin{eqnarray}\nonumber
\vec S_0(\zt) & = & \frac{1}{2\pi i}  \int_{-\infty}^0 d\tit \int_{-\infty}^{\infty} du \; e^{iu\zt} %\\ \nonumber %&&\times
 \mathbb{M}^T[\tilde \Omega^*(-\tit),-\tit,u] \; \vec a_{\mathrm{in}} (\tit),
\end{eqnarray}
giving the outcoming light mode
\begin{eqnarray}\nonumber \label{eq:aout}
\vec a_{\mathrm{out}}(\tit) & = &  \int_{-\infty}^0 d\tit'  \int_{-\infty}^\infty du \int_{-\infty}^{\infty} du'  
\left[ \frac{e^{i(u'-u)}-1}{-4\pi^2 i(u'-u)} \right] \\ \nonumber
&&  \times \mathbb{M}[\tilde \Omega(\tit),\tit,-u]  \mathbb{M}^T[\tilde \Omega^*(-\tit'),-\tit',u'] \; \vec a_{\mathrm{in}} (\tit').
\end{eqnarray}
Thus, the final expression for the efficiency of storage followed by backward retrieval reads
\begin{widetext}
\begin{eqnarray}
\eta_{\rm s+br} & = &  \int_0^\infty d\tit_x  \int_{-\infty}^0 d\tit  \int_{-\infty}^0 d\tit' 
\int_{-\infty}^\infty du \int_{-\infty}^\infty du' \int_{-\infty}^\infty d\tilde{u} \int_{-\infty}^\infty d\tilde{u}' 
 \vec a_{\mathrm{in}}^\dag(\tit) \; \left[ \frac{e^{-i(u'-u)}-1}{4\pi^2 i(u'-u)} \right]\\ \nonumber
&& \times \;\mathbb{M}^*[\tilde \Omega^*(-\tit),-\tit,u'] \;\mathbb{M}^\dag[\tilde \Omega(\tit_x),\tit_x,-u] 
 \mathbb{M}[\tilde \Omega(\tit_x),\tit_x,-\tilde{u}]  \; \mathbb{M}^T[\tilde \Omega^*(-\tit'),-\tit',\tilde{u}'] 
\; \left[ \frac{e^{i(\tilde{u}'-\tilde{u})}-1}{-4\pi^2 i(\tilde{u}'-\tilde{u})} \right] \; \vec a_{\mathrm{in}} (\tit').
\end{eqnarray}
\end{widetext}

In the position space and frequency picture, the relation between the incoming and outcoming light mode is
\begin{eqnarray}
\vec{a}_{\text{out}}(\omt) & = & \frac{1}{2\pi i} \int_{0}^{1}d\tilde{z} \int_{-i\cdot\infty}^{i\cdot\infty}d\omt'
\; \mathbb{K}[\tilde \Omega,\omt,\tilde{z}] \\ \nonumber
&& \times \mathbb{K}^{\text{T}}[\tilde \Omega^{*},\omt',\tilde{z}]\vec{a}_{\text{in}}(\omt'),
\end{eqnarray}
which leads to the final expression for the memory working in the backward direction
\begin{eqnarray} \nonumber
\eta_{\text{s+br}} & = & \frac{1}{8\pi^3 i} \int_{-i\cdot\infty}^{i\cdot\infty}d\omt \int_0^1 d \zt \int_0^1 d \zt'  
\int_{-i\cdot\infty}^{i\cdot\infty}d\nu \int_{-i\cdot\infty}^{i\cdot\infty}d\nu' \\
&& \times \vec a_{\text{in}}^\dag(\nu)  \mathbb{K}^*[\tilde \Omega^*, \nu, \zt]  \mathbb{K}^\dag[\tilde \Omega, \omt, \zt] \\ \nonumber
&& \times \mathbb{K}[\tilde \Omega, \omt, \zt']  \mathbb{K}^T[\tilde \Omega^*, \nu', \zt'] \vec a_{\text{in}}(\nu').
\end{eqnarray}
Again, we optimize the memory by looking at the eigenvalues and eigenvectors of the kernel matrix 
$\sim \mathbb{K}^*[\tilde \Omega^*, \nu, \zt]  \mathbb{K}^\dag[\tilde \Omega, \omt, \zt] 
\mathbb{K}[\tilde \Omega, \omt, \zt']  \mathbb{K}^T[\tilde \Omega^*, \nu', \zt']$
or $\sim \mathbb{M}^*[\tilde \Omega^*(-\tit),-\tit,u'] \mathbb{M}^\dag[\tilde \Omega(\tit_x),\tit_x,-u] 
 \mathbb{M}[\tilde \Omega(\tit_x),\tit_x,-\tilde{u}]$ $\mathbb{M}^T[\tilde \Omega^*(-\tit'),-\tit',\tilde{u}']$,
which gives us the efficiencies for the storage with backward read-out and the optimized incoming light modes, respectively.

%%%%%%%%%%%%%%%%%%%%%%%%%%%%%%%%%%%%%%%%%%
\section{Proof of the relation between the input and output light mode}\label{app:in_out}
%%%%%%%%%%%%%%%%%%%%%%%%%%%%%%%%%%%%%%%%%%

In this Appendix, we prove that the output light is the time reversed complex conjugate of the input mode
$\vec a_{\rm out}(\tit) = \vec a_{\rm in}^*(-\tit)$ 
for both backward and forward operating memories. 
We start with the forward memory and the beam splitter relation between the input and output light modes in the adiabatic limit
with a constant Rabi frequency $\tilde \Omega$ and make a substitution of the integral time variable $\tit' \rightarrow -\tit'$ leading to:
\begin{eqnarray}\nonumber
\vec a_{\mathrm{out}}(\tit) & = &\int_{-\infty}^\infty du  \int_0^{\infty} d\tit' \;  \frac{e^{iu}}{2\pi i}   \; \mathbb{M}[\tit,u] 
\mathbb{M}^T[\tit',u] \; \vec a_{\mathrm{in}} (-\tit').
\end{eqnarray}
By looking at the transpose of the matrix $\mathbb{D_{\rm f}}[\tit,\tit'] = \int_{-\infty}^\infty du \mathbb{M}[\tit,u] \mathbb{M}^T[\tit',u] 
=  \int_{-\infty}^\infty du \mathbb{Q}(u)e^{ \mathbb{N}(u)\tit}e^{ \mathbb{N}^T(u)\tit'} \mathbb{Q}^T(u)$ one can see that it is symmetric
$\mathbb{D_{\rm f}}[\tit,\tit'] = \mathbb{D_{\rm f}}^T[\tit',\tit]$.
Singular value decomposition of this matrix can be written as $\mathbb{D}[\tit,\tit'] = \sum_i |u_i \rangle \sqrt{\eta_i} \langle v_i |$
with the square roots of the corresponding efficiencies. The fact that this matrix is symmetric implies
that $|u_i\rangle = |v_i^*\rangle$, which means that the output light is the time reversed complex conjugate of the input light mode,
$\vec a_{\rm out}(\tit) = \vec a_{\rm in}^*(-\tit)$.
The transverse intensity shape of the output mode is the same and the phase is the complex conjugate of the input light mode.

Similarly for the backward operating memory, one can substitute $\tit' \rightarrow -\tit'$ and $u \rightarrow -u$
and write the beamsplitter relation in Eq.~(\ref{eq:aout}) as
\begin{eqnarray}\nonumber
\vec a_{\mathrm{out}}(\tit) & = &  \int_0^{\infty} d\tit'  \int_{-\infty}^\infty du \int_{-\infty}^{\infty} du'  
\left[ \frac{e^{i(u'+u)}-1}{-4\pi^2 i(u'+u)} \right] \\ \nonumber
&&  \times \mathbb{M}[\tit,u]  \mathbb{M}^T[\tit',u'] \; \vec a_{\mathrm{in}} (-\tit').
\end{eqnarray}
Again, we consider the transpose of the matrix  $\mathbb{D_{\rm b}}[\tit,\tit'] =  \int_{-\infty}^\infty du \int_{-\infty}^{\infty} du'  
\mathbb{M}[\tit,u] \mathbb{M}^T[\tit',u']  =  \int_{-\infty}^\infty du \int_{-\infty}^{\infty} du'    
\mathbb{Q}(u)e^{ \mathbb{N}(u)\tit}e^{ \mathbb{N}^T(u')\tit'} \mathbb{Q}^T(u')$,
which reads $\mathbb{D_{\rm b}}^T[\tit',\tit] =  \int_{-\infty}^\infty du \int_{-\infty}^{\infty} du'  
 \mathbb{Q}(u')e^{ \mathbb{N}(u')\tit}e^{ \mathbb{N}^T(u)\tit'} \mathbb{Q}^T(u)$.
Substituting $u\rightarrow u'$, we see that this matrix is symmetric, $\mathbb{D_{\rm b}}[\tit,\tit'] = \mathbb{D_{\rm b}}^T[\tit',\tit]$
so that also in the case of backward read-out we have the above symmetry between the input and output light modes.

%%%%%%%%%%%%%%%%%%%%%%%%%%%%%%%%%%%%%%%%%%
\section{Influence of the purity on the efficiencies}\label{app:purity}
%%%%%%%%%%%%%%%%%%%%%%%%%%%%%%%%%%%%%%%%%%

As we have shown in Sec.~\ref{sec:results}, the optimal input light modes are not necessarily pure,
however, the impurity is small and does not go above $4.5\%$ for the parameters considered here.
Experimentally, one would prefer to use pure light pulses to avoid the difficulty of changing the transverse profile in time.
Similarly, it is preferable to detect only the dominant Schmidt component of the corresponding output light. 
In the main text, we present efficiencies of the fully optimized pulses without constraints on light pulse purities.
In the present appendix, we estimate an upper bound to how much the experimental simplifications
of using separable modes $h_n(\vec r_\perp)$ and $a_n(\tit)$ will affect the efficiencies.

One can write the optimal input and output light modes as a sum of its dominant Schmidt component
$\vec{a}_{\rm in/out}^{\rm p}(\tit) = \vec{a}_{\rm in/out}^{\rm p*} g_{\rm in/out}(\tit)$
and the remaining impure part $\vec a_{\rm in/out}^{\rm \; ip}(\tit)$ (correlated in time $\tit$ and transverse position $\vec r_\perp$)
\begin{eqnarray}
\vec a_{\rm in}^{\rm \;opt}(\tit) & = & \sqrt{P_{\rm in}} \vec a_{\rm in}^{\rm \; p}(\tit) + \sqrt{1- P_{\rm in}} \vec a_{\rm in}^{\rm \; ip}(\tit), \\ \nonumber
\vec a_{\rm out}^{\rm \;opt}(\tit) & = & \sqrt{P_{\rm out}} \vec a_{\rm out}^{\rm \; p}(\tit) + \sqrt{1- P_{\rm out}} \vec a_{\rm out}^{\rm \; ip}(\tit),
\end{eqnarray}
respectively.
In the space of input light modes, we now consider the 2-dimensional
subspace $\mathbb{L}_{\text{in}}$ spanned by the ordered orthonormal
basis $\beta_{\text{in}}=\{\vec{a}_{\text{in}}^{\text{p}}(\tilde{t}),\vec{a}_{\text{in}}^{\text{ip}}(\tilde{t})\}$.
Obviously $\vec{a}_{\text{in}}^{\text{opt}}(\tilde{t})\in\mathbb{L}_{\text{in}}$
with the coordinate representation $\vec{e}_{1}=\left(\begin{array}{c}
\sqrt{P_{\text{in}}}\\
\sqrt{1-P_{\text{in}}}
\end{array}\right)$. We then define $\vec{a}_{\text{in}}^{\perp}(\tilde{t})\in\mathbb{L}_{\text{in}}$,
which is orthogonal to $\vec{a}_{\text{in}}^{\text{opt}}(\tilde{t})$,
as the mode with coordinate representation $\vec{e}_{2}=\left(\begin{array}{c}
-\sqrt{1-P_{\text{in}}}\\
\sqrt{P_{\text{in}}}
\end{array}\right)$. Now we consider the map $\mathbb{G}$ describing storage followed
by retrieval restricted to the domain $\mathbb{L}_{\text{in}}$ ($\mathbb{G}$
is an abbreviation for input-output relations of the type that appear in Eq.~(\ref{eq:ag})); 
denote the range of $\mathbb{G}$ by $\mathbb{L}_{\text{out}}$.
By definition, $\vec{a}_{\text{in}}^{\text{opt}}(\tilde{t})$ is an
eigenmode of $\mathbb{G}^{\dagger}\mathbb{G}$ with eigenvalue $\eta_{\text{max}}$
and we have $\sqrt{\eta_{\text{max}}}\vec{a}_{\text{out}}^{\text{opt}}(\tilde{t})=\mathbb{G}\vec{a}_{\text{in}}^{\text{opt}}(\tilde{t})$.
We define the normalized mode $\vec{a}_{\text{out}}^{\perp}(\tilde{t})$
by $\sqrt{\eta'}\vec{a}_{\text{out}}^{\perp}(\tilde{t})=\mathbb{G}\vec{a}_{\text{in}}^{\perp}(\tilde{t})$
and since $\vec{a}_{\text{in}}^{\perp}(\tilde{t})\perp\vec{a}_{\text{in}}^{\text{opt}}(\tilde{t})$, we have
\begin{eqnarray}\nonumber
\vec{a}_{\text{out}}^{\perp\dagger}(\tilde{t})\cdot\vec{a}_{\text{out}}^{\text{opt}}(\tilde{t}) 
& = &\frac{1}{\sqrt{\eta_{\text{max}}\eta'}}[\vec{a}_{\text{in}}^{\perp\dagger}(\tilde{t})\mathbb{G}^{\dagger}]
\cdot \left[\mathbb{G}\vec{a}_{\text{in}}^{\text{opt}}(\tilde{t})\right] \\\nonumber
& = &\frac{1}{\sqrt{\eta_{\text{max}}\eta'}}\vec{a}_{\text{in}}^{\perp\dagger}(\tilde{t})
\cdot[\mathbb{G}^{\dagger}\mathbb{G}\vec{a}_{\text{in}}^{\text{opt}}(\tilde{t})]\\\nonumber
& = &\sqrt{\frac{\eta_{\text{max}}}{\eta'}}\vec{a}_{\text{in}}^{\perp\dagger}(\tilde{t})\cdot\vec{a}_{\text{in}}^{\text{opt}}(\tilde{t})=0,
\end{eqnarray}
that is, $\vec{a}_{\text{out}}^{\perp}(\tilde{t})\perp\vec{a}_{\text{out}}^{\text{opt}}(\tilde{t})$;
here $\cdot$ signifies the inner product in Bessel mode
space and time. Next, we introduce an ordered orthonormal set $\beta_{\text{out}}$
of output light modes, $\beta_{\text{out}}=\{\vec{a}_{\text{out}}^{\text{p}}(\tilde{t}),\vec{a}_{\text{out}}^{\text{ip}}(\tilde{t}),\vec{b}(\tilde{t})\}$,
where $\vec{b}(\tilde{t})$ is a normalized mode that fulfills 
$c\vec{b}(\tilde{t})=\vec{a}_{\text{out}}^{\perp}(\tilde{t}) - 
[\vec{a}_{\text{out}}^{\text{p}\dagger}(\tilde{t})\cdot\vec{a}_{\text{out}}^{\perp}(\tilde{t})]
\vec{a}_{\text{out}}^{\text{p}}(\tilde{t}) - 
[\vec{a}_{\text{out}}^{\text{ip}\dagger}(\tilde{t})\cdot\vec{a}_{\text{out}}^{\perp}(\tilde{t})]
\vec{a}_{\text{out}}^{\text{ip}}(\tilde{t})$,
for some $c\in\mathbb{C}$. We have $\mathbb{L}_{\text{out}}\subseteq\text{span}(\beta_{\text{out}})$
and the coordinate representations w.r.t. $\beta_{\text{out}}$ of
$\vec{a}_{\text{out}}^{\text{opt}}(\tilde{t})$ and $\vec{a}_{\text{out}}^{\perp}(\tilde{t})$
are $\vec{f}_{1}=\left(\begin{array}{c}
\sqrt{P_{\text{out}}}\\
\sqrt{1-P_{\text{out}}}\\
0
\end{array}\right)$ and $\vec{f}_{2}=\frac{e^{i\phi}}{\sqrt{1+|c'|^{2}}}\left(\begin{array}{c}
-\sqrt{1-P_{\text{out}}}\\
\sqrt{P_{\text{out}}}\\
c'
\end{array}\right)$, respectively, for some $\phi\in\mathbb{R},c'\in\mathbb{C}$;
the ratio of the first two coordinates of $\vec{f}_{2}$ follows from
$\vec{a}_{\text{out}}^{\perp}(\tilde{t})\perp\vec{a}_{\text{out}}^{\text{opt}}(\tilde{t})$.

We can now deduce a lower bound for the efficiency $\eta_{\text{pure}}$
of storing the pure mode $\vec{a}_{\text{in}}^{\text{p}}(\tilde{t})$
followed by retrieval where the output light is projected into the
spatial profile $\vec{a}_{\text{out}}^{\text{p}}$ of the pure mode
$\vec{a}_{\text{out}}^{\text{p}}(\tilde{t})=\vec{a}_{\text{out}}^{\text{p}}g_{\text{out}}^{\text{p}}(\tilde{t})$.
Note, however, that in general $\vec{a}_{\text{out}}^{\text{p}}$
is not the spatial profile with the largest integrated overlap with
the output light corresponding to $\vec{a}_{\text{in}}^{\text{p}}(\tilde{t})$.
In the coordinate space of $\mathbb{L}_{\text{in}}$, $\vec{a}_{\text{in}}^{\text{p}}(\tilde{t})$
is represented by
$\left(\begin{array}{c}
1\\ 0 \end{array}\right)=\sqrt{P_{\text{in}}}\vec{e}_{1}-\sqrt{1-P_{\text{in}}}\vec{e}_{2}$, whereby
\begin{eqnarray}\nonumber
\eta_{\text{pure}} & = & |\left(\begin{array}{ccc}
1 & 0 & 0\end{array}\right)[\mathbb{G}]_{\beta_{\text{in}}}^{\beta_{\text{out}}}\left(\begin{array}{c}
1\\ 0 \end{array}\right)|^{2}\\ \nonumber
& = & |\left(\begin{array}{ccc}
1 & 0 & 0\end{array}\right)(\sqrt{\eta_{\text{max}}}\sqrt{P_{\text{in}}}\vec{f}_{1}-\sqrt{\eta'}\sqrt{1-P_{\text{in}}}\vec{f}_{2})|^{2}\\\nonumber
& = & |\sqrt{\eta_{\text{max}}}\sqrt{P_{\text{in}}}\sqrt{P_{\text{out}}}\\ \nonumber
&& +\sqrt{\eta'}\sqrt{1-P_{\text{in}}}\sqrt{1-P_{\text{out}}}\frac{e^{i\phi}}{\sqrt{1+|c'|^{2}}}|^{2}.
\end{eqnarray}
Assuming that the largest eigenvalue $\eta_{\text{max}}$ of $\mathbb{G}^{\dagger}\mathbb{G}$
is non-degenerate (which seems to be generally true from numerical
calculations~\cite{Grodecka-Grad2011}), we have from $\vec{a}_{\text{in}}^{\perp}(\tilde{t})\perp\vec{a}_{\text{in}}^{\text{opt}}(\tilde{t})$
that $\eta'\leq\eta_{2}$, where $\eta_{2}$ is the second largest
eigenvalue. However, since only the values of $\eta_{\text{max}}$
are presented in the present paper, we will assume $\eta'=\eta_{\text{max}}$
resulting in a smaller lower bound. Maximal destructive interference
will occur for $\phi=\pi$ and $c'=0$; making these substitutions
along with $P_{\text{out}}=P_{\text{in}}$ (as follows from the result
of Appendix~\ref{app:in_out}), we obtain
\[
\eta_{\text{pure}}\geq\eta_{\text{max}}(1-2P_{\text{in}})^{2}=\eta_{\text{max}}(1-2\varepsilon)^{2}\approx\eta_{\text{max}}(1-4\varepsilon),
\]
where $\epsilon=1-P_{\text{in}}$ is the impurity.
The efficiency of the full optimal mode in the worst case of $P = 0.9581$ 
for forward retrieval, $m=0$,  $d_0=200$, and $F=0.02$ can at most be reduced by 
$\sim 16.76\%$ (from $\eta = 0.8049$ to $\eta_{\rm pure} = 0.67$).
\end{appendix}

\bibliographystyle{prsty}
\bibliography{longQM}

\begin{thebibliography}{10}

\bibitem{Hammerer2010}
K. Hammerer, A.~S. S{\o}rensen, and E.~S. Polzik, Rev. Mod. Phys. {\bf 82},
  1041  (2010).

\bibitem{Sangouard2011}
N. Sangouard, C. Simon, H. de~Riedmatten, and N. Gisin, Rev. Mod. Phys. {\bf
  83},  33  (2011).

\bibitem{duan2001}
L.~M. Duan, M.~D. Lukin, J.~I. Cirac, and P. Zoller, Nature {\bf 414},  413
  (2001).

\bibitem{kimble2008}
H.~J. Kimble, Nature {\bf 453},  1023  (2008).

\bibitem{Gorshkov2007}
A.~V. Gorshkov, A. Andr\'{e}, M.~D. Lukin, and A.~S. S{\o}rensen, Phys. Rev. A
  {\bf 76},  033806  (2007).

\bibitem{Boozer2007}
A. Boozer, A. Boca, R. Miller, T. Northup, and H.~J. Kimble, Phys. Rev. Lett.
  {\bf 98},  193601  (2007).

\bibitem{Specht2011}
H.~P. Specht, C. N\"{o}lleke, A. Reiserer, M. Uphoff, E. Figueroa, S. Ritter,
  and G. Rempe, Nature {\bf 473},  190  (2011).

\bibitem{Sorensen2008}
M.~W. S{\o}rensen and A.~S. S{\o}rensen, Phys. Rev. A {\bf 77},  013826
  (2008).

\bibitem{Nunn2008}
J. Nunn, K. Reim, K.~C. Lee, V.~O. Lorenz, B.~J. Sussman, I.~A. Walmsley, and
  D. Jaksch, Phys. Rev. Lett. {\bf 101},  260502  (2008).

\bibitem{Julsgaard2004}
B. Julsgaard, J. Sherson, J.~I. Cirac, J. Fiur\'{a}sek, and E.~S. Polzik,
  Nature {\bf 432},  482  (2004).

\bibitem{Novikova2007}
I. Novikova, A.~V. Gorshkov, D.~F. Phillips, A.~S. S{\o}rensen, M.~D. Lukin,
  and R.~L. Walsworth, Phys. Rev. Lett. {\bf 98},  243602  (2007).

\bibitem{Vudyasetu2008}
P.~K. Vudyasetu, R.~M. Camacho, and J.~C. Howell, Phys. Rev. Lett. {\bf 100},
  123903  (2008).

\bibitem{Firstenberg2009}
O. Firstenberg, P. London, M. Shuker, A. Ron, and N. Davidson, Nat. Phys. {\bf
  5},  665  (2009).

\bibitem{Choi2010}
K.~S. Choi, A. Goban, S.~B. Papp, S.~J. van Enk, and H.~J. Kimble, Nature {\bf
  468},  412  (2010).

\bibitem{Reim2010}
K.~F. Reim, J. Nunn, V.~O. Lorenz, B.~J. Sussman, K.~C. Lee, N.~K. Langford, D.
  Jaksch, and I.~A. Walmsley, Nat. Photon. {\bf 4},  218  (2010).

\bibitem{Hosseini2011}
M. Hosseini, B.~M. Sparkes, G. Campbell, P.~K. Lam, and B.~C. Buchler, Nat.
  Commun. {\bf 2},  174  (2011).

\bibitem{Afzelius2010}
M. Afzelius, I. Usmani, A. Amari, B. Lauritzen, A. Walther, C. Simon, N.
  Sangouard, J. Min\'{a}ř, H. de~Riedmatten, N. Gisin, and S. Kr\"{o}ll, Phys.
  Rev. Lett. {\bf 104},  040503  (2010).

\bibitem{Bonarota2011}
M. Bonarota, J.-L. {Le Gou\"{e}t}, and T. Chaneli\`{e}re, New J. Phys. {\bf
  13},  013013  (2011).

\bibitem{Saglamyurek2011}
E. Saglamyurek, N. Sinclair, J. Jin, J.~A. Slater, D. Oblak, F. Bussi\`{e}res,
  M. George, R. Ricken, W. Sohler, and W. Tittel, Nature {\bf 469},  512
  (2011).

\bibitem{Sorensen2009}
M.~W. S{\o}rensen and A.~S. S{\o}rensen, Phys. Rev. A {\bf 80},  033804
  (2009).

\bibitem{Porras2008}
D. Porras and J. Cirac, Phys. Rev. A {\bf 78},  053816  (2008).

\bibitem{Pedersen2009}
L.~H. Pedersen and K. M{\o}lmer, Phys. Rev. A {\bf 79},  012320  (2009).

\bibitem{Gorshkov2007a}
A.~V. Gorshkov, A. Andr\'{e}, M.~D. Lukin, and A.~S. S{\o}rensen, Phys. Rev. A
  {\bf 76},  033805  (2007).

\bibitem{Gorshkov2007c}
A.~V. Gorshkov, A. Andr\'{e}, M. Fleischhauer, A.~S. S{\o}rensen, and M.~D.
  Lukin, Phys. Rev. Lett. {\bf 98},  123601  (2007).

\bibitem{Liu2001}
C. Liu, Z. Dutton, C.~H. Behroozi, and L.~V. Hau, Nature {\bf 409},  490
  (2001).

\bibitem{Kozhekin2000}
A. Kozhekin, K. M{\o}lmer, and E. Polzik, Phys. Rev. A {\bf 62},  033809
  (2000).

\bibitem{Moiseev2001}
S. Moiseev and S. Kr\"{o}ll, Phys. Rev. Lett. {\bf 87},  173601  (2001).

\bibitem{Surmacz2008}
K. Surmacz, J. Nunn, K. Reim, K.~C. Lee, V.~O. Lorenz, B. Sussman, I.~A.
  Walmsley, and D. Jaksch, Phys. Rev. A {\bf 78},  033806  (2008).

\bibitem{Pohl2010}
T. Pohl, E. Demler, and M.~D. Lukin, Phys. Rev. Lett. {\bf 104},  043002
  (2010).

\bibitem{Andre2005}
A. Andre, Ph.D. thesis, 2005.

\bibitem{Grodecka-Grad2011}
A. Grodecka-Grad, E. Zeuthen, and A.~S. S{\o}rensen (unpublished).

\bibitem{bartels1972}
R.~H. Bartels and G.~W. Stewart, Commun. ACM {\bf 15},  820  (1972).

\end{thebibliography}

\end{document}